\documentclass[useAMS,usenatbib]{mn2e} 

\usepackage[english]{babel}
\usepackage{journals}
\usepackage[pdftex]{graphicx,color} 
\usepackage[latin1]{inputenc}
\usepackage{graphics} 
\usepackage{amsfonts} 
\usepackage{amsmath}
\usepackage{multicol} 
\usepackage{layout} 
\usepackage{amssymb}
\usepackage[a4paper,colorlinks=true,pdfstartview=FitV,linkcolor=red,citecolor=blue,urlcolor=magenta]{hyperref}
\usepackage{layouts}
\usepackage{multirow}
\usepackage{float}
\usepackage[caption = false]{subfig}
\usepackage{tabularx}
\usepackage{siunitx}

\title[SHMR of satellite galaxies]{
Stellar-to-halo mass relation of cluster galaxies
}
\author[Niemiec et al.]
	{\parbox{\textwidth}{
	Anna Niemiec$^{1}$\thanks{E-mail: \href{mailto:anna.niemiec@lam.fr} {anna.niemiec@lam.fr}}, 
	Eric Jullo$^{1}$, 
	Marceau Limousin$^{1}$,
	Carlo Giocoli$^{1,2}$, 
	Thomas Erben$^{3}$,
	Hendrik Hildebrant$^{3}$,
	Jean-Paul Kneib$^{4}$,
	Alexie Leauthaud$^{5}$,
	Martin Makler$^{6}$,
	Bruno Moraes$^{7}$,
	Maria E. S. Pereira$^{6}$,
	Huanyuan Shan$^{3,4}$,
	Eduardo Rozo$^{8}$,
	Eli Rykoff$^{9}$ and
	Ludovic Van Waerbeke$^{10}$
	\\ \\ 
	}
	\vspace*{3pt}\\
$^{1}$ Aix Marseille Universit\'e, CNRS, LAM (Laboratoire d'Astrophysique de Marseille) UMR 7326, 13388, Marseille, France \\
$^{2}$ Dipartimento di Fisica e Astronomia, Alma Mater Studiorum Universit\`a di Bologna, viale Berti Pichat, 6/2, 40127 Bologna, Italy \\
$^{3}$ Argelander-Institut f\"ur Astronomie, Auf dem H\"ugel 71, 53121 Bonn, Germany \\
$^{4}$ Laboratoire d'astrophysique (LASTRO), Ecole Polytechnique F\'ed\'erale de Lausanne (EPFL), Observatoire de Sauverny, CH-1290 Versoix, Switzerland \\
$^{5}$ Department of Astronomy and Astrophysics, University of California, Santa Cruz, 1156 High Street, Santa Cruz, CA 95064 USA \\
$^{6}$ Centro Brasileiro de Pesquisas F\'isicas, Rua Dr Xavier Sigaud 150, CEP 22290-180, Rio de Janeiro, RJ, Brazil \\
$^{7}$ Dept. of Physics and Astronomy, University College London, London, WC1E 6BT, UK \\
$^{8}$ Department of Physics, University of Arizona, 1118 E. Fourth St., Tucson, AZ 85721, U.S.A. \\
$^{9}$ SLAC National Accelerator Laboratory, Menlo Park, CA 94025, U.S.A. \\
$^{10}$ University of British Columbia, Department of Physics and Astronomy, 6224 Agricultural road, V6T 1Z1, Vancouver, Canada \\
   }
   
\begin{document}
\date{}
\maketitle
\label{firstpage}
\pagerange{\pageref{firstpage}--\pageref{lastpage}} \pubyear{2017}

\begin{abstract}


In the formation of galaxy groups and clusters, the dark matter haloes containing satellite galaxies are expected to be tidally stripped in gravitational interactions with the host. We use galaxy-galaxy weak lensing to measure the average mass of dark matter haloes of satellite galaxies as a function of projected distance to the centre of the host, since stripping is expected to be greater for satellites closer to the centre of the cluster. We further classify the satellites according to their stellar mass: assuming that the stellar component of the galaxy is less disrupted by tidal stripping, stellar mass can be used as a proxy of the infall mass. We study the stellar to halo mass relation of satellites as a function of the cluster-centric distance to measure tidal stripping.
We use the shear catalogues of the DES science verification archive, the CFHTLenS and the CFHT Stripe 82 surveys, and we select satellites from the redMaPPer catalogue of clusters.
For galaxies located in the outskirts of clusters, we find a stellar to halo mass relation in good agreement with the theoretical expectations from \citet{moster2013} for central galaxies. In the centre of the cluster, we find that this relation is shifted to smaller halo mass for a given stellar mass. We interpret this finding as further evidence for tidal stripping of dark matter haloes in high density environments.

\end{abstract}


\section{Introduction}

Galaxy clusters are large structures in the Universe, composed of tens to hundreds of galaxies bound by gravity. In the hierarchical formation model, they are formed and grow by accretion of smaller groups or isolated galaxies. In this scenario, understanding how these accreted galaxies interact with the very dense cluster environnement is an important step towards explaining the global picture of galaxy evolution and structure formation \citep{kauffmann1999, kauffmann1999b}.

Numerous studies have been performed on the properties of subhaloes in numerical simulations \citep[see for example][]{ghigna1998, delucia2004, gao2004, contini2012, vandenbosch2015}. 
They predict that during infall, subhaloes are subject to the tidal forces of their hosts, which strip from them part of their dark matter \citep{hayashi2008, giocoli2008}.  Subhaloes which have been accreted earlier have experienced tidal stripping from the host for a longer time, and have thus lost a higher fraction of their initial mass. See also \citet{gao2004, vandenbosch2005b, limousin2009, giocoli2010}. 

In addition, there is a correlation between the distance of the subhalo to the cluster centre and the time since accretion, with subhaloes accreted earlier residing on more tightly bound orbits. This effect is due on one hand to the inside-out assembly of dark matter haloes (i.e. at higher accretion redshift the host halo was smaller, leading to satellites being accreted at smaller cluster-centric distance compared to later redshifts \citep{tormen1998}), and on the other hand to dynamical friction that slows subhaloes down and make them sink into the centre of the host as time passes \citep{gao2004}.
So observationally, to study the evolution of subhaloes during infall, we can use the distance from the satellite to the centre of the cluster as an indicator of the time since the accretion of the satellite. 
In observations, the distances that are measured are projected along the line of sight, but \citet{vandenbosch2015} shows the correlation between this projected distance and the accretion redshift is still very strong, although weaker than for the 3-dimensional distance.
 
Note that \citet{vandenbosch2015} shows that the segregation is much stronger if the present subhalo mass is normalized by its mass at accretion. Indeed looking at the global satellite population, tidal stripping and dynamical friction have opposite effects on the radial distribution of subhalo masses in the cluster. While stripping tends to reduce more strongly the mass of satellites close to the centre, the satellites which are bigger at accretion are subjected to stronger dynamical friction and thus more driven to the centre.

Since in observations the mass at accretion is not an observable, we use as a proxy the stellar mass. As the host tidal forces strip preferentially the outer part of the subhaloes, the baryonic part at their centre is not significantly disturbed. While it is also possible that stellar mass is created during infall, most galaxies have their star formation quenched at accretion \citep{zu2015, zu2016}.   The stellar mass is thus well correlated to the subhalo mass at accretion \citep{vale2006, cooray2006, behroozi2010, smith2016}.

In summary, we measure $M_{\rm sat}^{\rm DM}/M_{\rm sat}^{\rm star}$ as a function of the projected distance from the satellite to the cluster centre, to see how the dark matter halo of the satellites is affected by the tidal stripping of the host throughout infall.
   
A well-established tool to measure the total projected mass of a galaxy, including the dark matter halo, is gravitational lensing. The light coming from background sources is deviated when it passes by a massive (lens) object, and the amplitude of the deviation depends on the total mass of the lens object. Lensing measurements of the mass of an object are independent of its nature (baryons, dark matter, etc.) and its state (equilibrium, etc.).

The first measure of the mass of subhaloes in a cluster was performed by \citet{natarajan1998} in the cluster AC114, measuring the perturbation by subhaloes of the shear distribution of background sources. In a later work on the cluster Cl0024+16 using HST images, \citet{natarajan2009} present the first indication for tidal stripping on the dark matter haloes of satellite galaxies.
Later, \cite{okabe2014} performed a similar analysis for the Coma cluster using data from the Subaru telescope.
 
Alternatively, the dark matter mass of satellites can be measured in a statistical way using galaxy-galaxy weak lensing over a large sample of galaxies. 
It consists of the measurement of the average tangential shear in the shape of background galaxies  in circular bins centred on a lens object, in order to measure the mass of the lens. In galaxy-galaxy lensing the lens is the dark matter halo of a galaxy, and so has a relatively low mass: the shear induced by a single galactic halo is very low, so to obtain a measurable lensing signal the shear induced by numerous lenses needs to be stacked together \citep{brainerd1996}. 

\citet{yang2006} first suggested to apply the galaxy-galaxy lensing method to measure the mass of subhaloes in clusters, and \citet{li2013b} first measured the mass of satellite galaxies in groups using data from SDSS and the CFHT Stripe 82 (CS82) survey.
Using this method, \citet{gillis2013} measured the total mass of galaxies in low and high density environments, arguing that the galaxies in high  density environment were subject to tidal stripping, compared to the ones in low density environment. In the $100\deg^2$ overlap between the DR2 of the KiDS lensing survey and the GAMA spectroscopic survey \cite{sifon2015} repeat the analysis but the statistical errors prevent them from measuring tidal stripping of dark matter for galaxies in clusters with a mass $M_{\rm host} > 10^{13}h^{-1}M_{\odot}$. Finally, in the $170\deg^2$ of the CFHT Stripe 82 survey, \cite{li2015} measured the mass of the dark matter haloes and the evolution of the mass-to-light ratio for satellite galaxies as a function of their distances to the centres of the redMaPPer clusters, and obtain a significant signal consistent with tidal stripping. Finally, \citet{vanuitert2016} measured the stellar-to-halo mass relation for galaxies from the GAMA survey, comparing central to satellite galaxies, and found no significant difference.
 
We extend these works by considering the stellar mass distribution of the satellites, and further splitting our galaxy samples according to it. As mentioned above, it appears from simulations that for satellites, the quantity which is more segregated  with respect to the distance to the centre of clusters is the present subhalo mass \textit{normalized by the mass at accretion}. We use the stellar mass as an observational proxy for the mass at accretion.

This paper is organized as follows: in Section \ref{sec:methodology}, we present the lensing method and the halo model we use to fit the data. In Section \ref{sec:data}, we present our source and lens catalogues, then in Section \ref{sec:results} we show the results of the analysis, and we discuss them in Section \ref{sec:discussion}. We conclude in Section \ref{sec:conclusion}.

We assume a WMAP7 cosmology \citep{komatsu2011}, with $(\Omega_{\rm M}, \Omega_{\Lambda}, h, \sigma_8, w) = (0.27,0.73,0.70,0.81,-1)$. When relevant, the dependence on $h$ is clearly stated.

\section{Methodology}
\label{sec:methodology}

	\subsection{Lensing}
	
In the weak lensing regime, the distortion induced on the image of source galaxies by a single lens galaxy is so weak that it is too tiny to be detected: the observed shape of a background source is the sum of its intrinsic ellipticity and of the lensing shear, and the shear represents only a few per cent of the total ellipticity. However, by stacking the measurements of many individual lensing signals together, the average tangential shear generated by a sample of lenses can be detected.

We measure the galaxy-galaxy lensing observable, the excess surface mass density profile $\Delta\Sigma(R)$  in comoving units as a function of the distance to the (stacked) lenses, and this quantity is related to the tangential shear $\gamma_t$ by:
\begin{equation}
	 \Delta\Sigma(R) = \Sigma_{\rm crit}\gamma_{\rm t}(R) = \bar{\Sigma}(<R) - \bar{\Sigma}(R) \mathrm{,}
\end{equation}
where $\bar{\Sigma}(R)$ is the mean surface density at a projected distance $R$ from the lens centre, $\bar{\Sigma}(<R)$ is the mean surface density in a disk of radius $R$ centred on the lens, and $\Sigma_{\rm crit}$ is the critical surface density in comoving units. It is defined as
\begin{equation}
	\Sigma_{\rm crit} = \frac{c^2}{4\pi G}\frac{D_{\rm s}}{D_{\rm l}D_{\rm ls}}\frac{1}{(1 + z_{\rm l})^2}
\end{equation}
where $D_{\rm l}$ and $D_{\rm s}$ are the angular diameter distances respectively to the lens and to the source, and $D_{\rm ls}$ is the angular diameter distance between the lens and the source. $z_{\rm l}$ is the redshift of the lens.

	\subsection{Halo Model}
	\label{sec:model}
	
The lensing signal around the satellite galaxies is described by a four-term halo model \citep[e.g.][]{cooray&sheth2002, sheth2003, giocoli2010, gillis2013}:
\begin{equation}
	\Delta\Sigma = \Delta\Sigma_{\rm star} + \Delta\Sigma_{\rm sat} + f_{\rm sat}\Delta\Sigma_{\rm host}  +\Delta\Sigma_{\rm 2h} \mathrm{.}
\end{equation}

Here $\Delta\Sigma_{\rm star}$ is the baryonic component of the satellites,  $\Delta\Sigma_{\rm sat}$ corresponds to the contribution of the satellite dark matter (ie the subhalo), $\Delta\Sigma_{\rm host}$ is the contribution of the host (cluster) haloes of the satellites, and $\Delta\Sigma_{\rm 2h}$ is the two-halo term, produced by the neighbouring haloes. The factor $f_{\rm sat}$ represents the fraction of satellites in the lens sample, and we fix $f_{\rm sat} = 1$ as we study the lensing signal produced by a theoretically pure satellite sample. We note that in the case $f_{\rm sat} \neq 1$ the term $\Delta\Sigma_{\rm sat}$ should actually be written $f_{\rm sat}\Delta\Sigma_{\rm sat} + (1 - f_{\rm sat})\Delta\Sigma_{\rm cent}$, with a contribution from both satellite (in subhaloes) and central galaxies (in haloes). We write only $\Delta\Sigma_{\rm sat}$ for clarity. We do not take into account a central baryonic component in the host clusters.

 The model is expressed in comoving units. We describe the different terms in this section.

		\subsubsection{The stellar component}
		
We consider the baryonic component of the satellite galaxy mass to be a point source, with mass $M_*$ equal to the median stellar mass of the sample. Thus:
\begin{equation}
	\Delta\Sigma_{\rm star}(R) = \frac{M_{*}}{R^2}\mathrm{.}
\end{equation}

		\subsubsection{The satellite term}

The galaxies are assumed to live in dark matter haloes, characterized by a NFW density profile \citep{NFW1996}. In the case of satellite galaxies, we use a smoothly truncated NFW profile, whose spatial density distribution is defined by \citep{baltz2009}:
\begin{equation}
	\rho(r) = \frac{\rho_{\rm crit}\delta_{\rm c}}{(r/r_{\rm s})(1+r/r_{\rm s})^2(1+(r/r_{\rm t})^2)} \mathrm{,}
\end{equation}
where $r_{\rm t}$ is the truncation parameter which ensures that the total NFW mass does not diverge. We fix it at $r_{\rm t} = 2r_{200}$ \citep{hilbert2010, oguri2011}, a value for which the profile hardly deviates from a standard NFW inside the virial radius. This profile has two free parameters: the halo mass $M_{200}$ and the concentration $c=r_{200}/r_{\rm s}$. The virial radius $r_{200}$ defines a sphere with a density 200 times higher than the critical density of the Universe, which gives
\begin{equation}
	\rho_{\rm crit} = \frac{3}{800\pi}\frac{M_{200}}{r_{200}^3} \mathrm{,}
\end{equation}
and $\delta_{\rm c}$ can be expressed as
\begin{equation}
	\delta_{\rm c} = \frac{200}{3}\frac{c^3}{\ln(1 + c) - \frac{c}{1 + c}} \mathrm{.}
\end{equation}

To reduce the number of free parameters, we use a mass-concentration relation, defined at $z=0$ in \cite{neto2007}:
\begin{equation}
	c_0 = 4.67 \times \left( \frac{M_{200}}{10^{14}h^{-1}M_{\odot}} \right)^{-0.11}
\end{equation}
and add the redshift dependence $c = c_0/(1+z)$. To verify the influence of this choice of a mass-concentration relation, we compare $\Delta\Sigma_{\rm sat}$ computed using different mass-concentration relations \citep{neto2007, duffy2008, dutton2014, shan2017} at $M_{sat}=10^{12}h^{-1}M_{\odot}$, and find a maximum deviation in the amplitude of the signal of $\sim 1\%$, which is negligible compared to the error bars on our measurements.

		\subsubsection{The host halo term}

\begin{figure}
	\centering
	\includegraphics{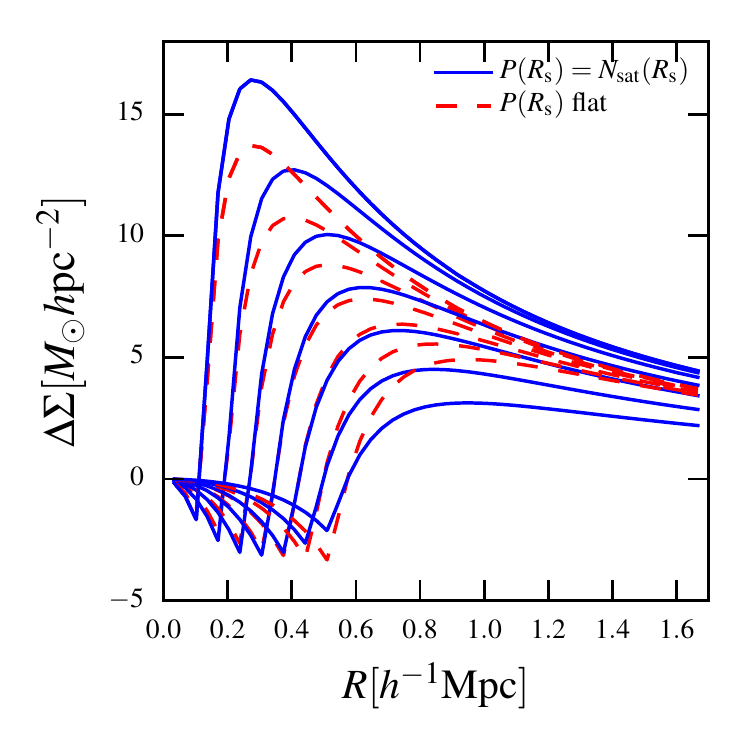}
	\caption{ Lensing signal produced by host haloes at different distances from the satellite $R_{\rm s} \in [0.12; 0.52]h^{-1}\si{Mpc}$. The dashed red curves represent hosts which all have the same weight, whereas for the blue lines hosts are weighted by $P(R_{\rm s}) = N_{\rm sat}(R_{\rm s})$. The total host term is a continuous sum of weighted single halo terms, as in equation \ref{eq_host}. For this plot we fixed $M_{\rm host}$ at $10^{14}h^{-1}M_{\odot}$ and use $N_{\rm sat}(R_{\rm s})$ from the satellites with $10.5 < \log(M_*/M_{\odot}) < 11$ and $0.1 < R_{\rm s}/h^{-1}\si{Mpc} < 0.55$.
          }
	\label{fig:hosts}
\end{figure}

As we measure the lensing signal of satellite galaxies in clusters, an important component of the total signal is the one induced by the host clusters.
We stack the lensing signal for satellites located at different projected distances to the centre of their cluster. The mean host halo signal is expressed as
\begin{equation} \label{eq_host}
	\Delta\Sigma_{\rm host}(R, M_{\rm host}) =  \frac{\int_{r_1}^{r_2} \Delta\Sigma_{\rm 1host}(R,R_{\rm s}, M_{\rm host})P(R_{\rm s})\operatorname{d}\!R_{\rm s}}{\int_{r_1}^{r_2} P(R_{\rm s})\operatorname{d}\!R_{\rm s}}  \mathrm{,}
\end{equation}
where $\Delta\Sigma_{\rm 1host}(R,R_{\rm s}, M_{\rm host})$ is the mean contribution of one host with mass $M_{\rm host}$ located at a projected distance $R_{\rm s}$ from the satellite, and $P(R_{\rm s})$ is the probability for a satellite to be located at a distance $R_{\rm s}$ from the centre of its host cluster. Therefore the term $\Delta\Sigma_{\rm host}(R, M_{\rm host})$ is the mean contribution of a host located at a distance $R_{\rm s}$ ranging between $r_1$ and $r_2$, and weighted by $P(R_{\rm s})$ which is described later. Figure \ref{fig:hosts} shows a sample of lensing signals produced by a host located at different distances to the satellite, both with and without the weighting. We assume that all the host haloes have the same mass, and we measure this average mass $M_{\rm host}$.

The contribution of one host halo located at a distance $R_{\rm s}$ from its satellite is defined as
\begin{equation}
	\Delta\Sigma_{\rm 1host}(R,R_{\rm s}) = \bar{\Sigma}_{\rm 1host}(<R, R_{\rm s}) - \bar{\Sigma}_{\rm 1host}(R,R_{\rm s})  \\
\end{equation}
\begin{multline}
	\Delta\Sigma_{\rm 1host}(R,R_{\rm s})= \frac{1}{\pi R^2} \int_0^R R' \int_0^{2\pi} \Sigma_{\rm 1host}(R_{\rm g}')\operatorname{d}\!\theta \operatorname{d}\!R' \\
		- \frac{1}{2\pi} \int_0^{2\pi} \Sigma_{\rm 1host}(R_{\rm g})\operatorname{d}\!\theta
\end{multline}
where $\Sigma_{\rm 1host}(R_{\rm g})$ is the projected surface density of the host halo, a NFW halo of mass $M_{\rm host}$, measured at a distance $R_{\rm g} = |\vec{R} - \vec{R_{\rm s}}| = \sqrt{R^2+R_{\rm s}^2-2RR_{\rm s}\cos{\theta}}$ from the host centre, where $\theta$ is the angle between the vector joining the centre of the satellite to the centre of the host, and the vector joining the centre of the satellite to the point of measurement. As we average the signal in circles or disks centred on the satellite, we integrate over $\theta$.

\begin{figure}
	\centering
	\includegraphics{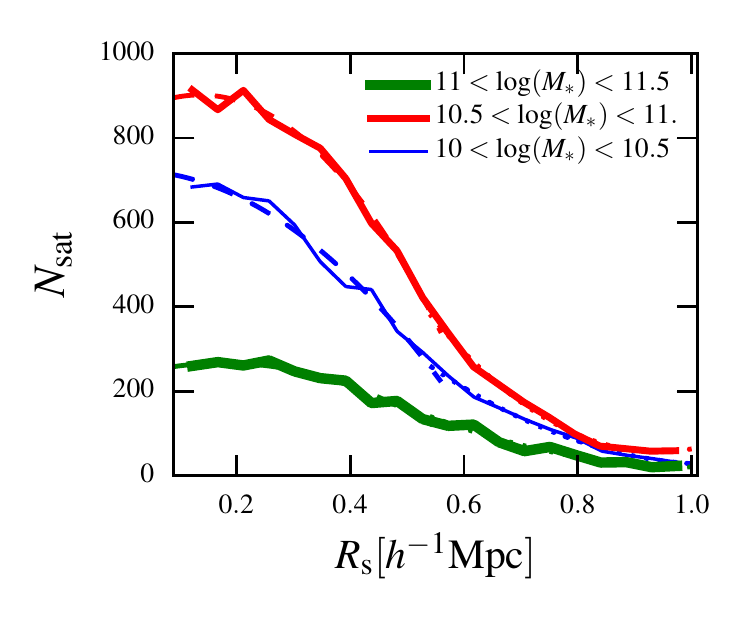}
	\caption{ Satellite spatial distribution in three bins in stellar mass: $10 < \log (M_*/M_{\odot}) < 10.5$, $10.5 < \log (M_*/M_{\odot}) < 11$ and $11 < \log (M_*/M_{\odot}) < 11.5$. The dashed and dotted curves are the best-fit second order polynomials in each cluster-centric bin.
          }
	\label{fig:sat_distrib}
\end{figure}

To obtain the probability function $P(R_{\rm s})$, we use the distribution of satellites in the data, stacked for all the clusters, in each bin of cluster-centric distance and stellar mass, ie. $P(R_{\rm s}) = N_{\rm sat}(R_{\rm s})$. In order to smooth the obtained distribution, for each stellar mass bin we fit the distribution with two second order polynomials, one for each cluster-centric bin. The $N_{\rm sat}$ distributions and the best-fit curves are shown in figure \ref{fig:sat_distrib}.

\begin{figure}
	\centering
	\includegraphics{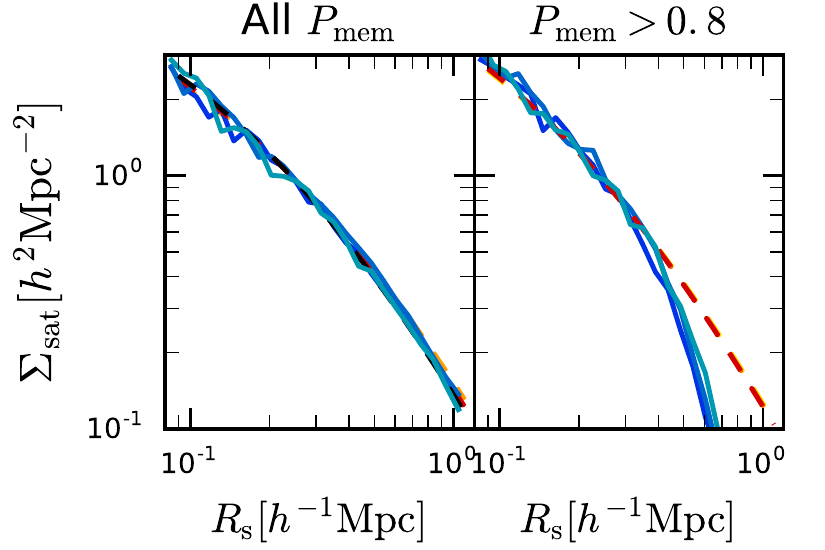}
	\caption{Surface number density of satellites in the three stellar mass bins. The continuous lines represent the stacked measurements, where each satellite is weighted by its probability of membership  to the host (see \ref{sec:lens_selection}). The left plot presents the measurements for all the satellites, while on the right plot the measurement is made for our fiducial sample, which contains only the satellites with $P_{\rm mem} > 0.8$. The dashed lines are the corresponding best-fit NFW profiles.
          }
	\label{fig:sat_density}
\end{figure}

\paragraph*{Satellite concentration.} Alternatively, we fit the surface number density profile of the satellites with a NFW profile to measure their concentration. We plot the surface number density profiles $\Sigma_{\rm sat}$, for each of the three stellar mass bins on figure \ref{fig:sat_density}. Each galaxy is weighted by the probability that it is a true member of its host $P_{\rm mem}$. The left panel of the figure presents the measurement for the whole satellite sample, while the right panel plot is for our fiducial sample which contains only the galaxies with $P_{\rm mem} > 0.8$ (see section \ref{sec:lens_selection}). For each profile we fit a projected NFW profile. We normalize the measurement and the model to have the same amplitude around the scale radius.

Once normalized, the profiles for galaxies in the different stellar mass bin are similar to each other, and the best-fit concentrations we obtain have very close values between the different bins. From low to high stellar mass bins, we obtain concentrations of 3, 3.4 and 3.7 for the complete satellite sample, and 4.3, 4.4 and 4.5 for the sample with $P_{\rm mem} > 0.8$. 

While the full satellite sample distribution is well described by a NFW density profile, the sample with $P_{\rm mem} > 0.8$ is steeper, which is why we fit the satellite distribution with polynomials in the model.

We note that if we do not weight the galaxies by their membership probability, the full sample gives a flatter distribution than the NFW profile, and the $P_{\rm mem}$ selected sample remains steeper.

		\subsubsection{The two-halo term}
		
On large scales the lensing signal is dominated by neighbouring mass concentrations, e.g., nearby haloes and filaments. The contribution of the neighbours is accounted for by the two-halo term, defined as:

\begin{equation}
	\Delta\Sigma_{\rm 2h} (R) =   \bar{\Sigma}_{\rm 2h} (< R) - \bar{\Sigma}_{\rm 2h} (R) 
\end{equation}
with
\begin{equation}
	\bar{\Sigma}_{\rm 2h} ( R) =  2\rho_{\rm c,0}\Omega_{\rm m,0}\int_0^{\infty} \xi_{\rm gm}^{\rm 2h}(\sqrt{R^2 + \chi^2})\operatorname{d}\!\chi
\end{equation}
and
\begin{equation}
	\bar{\Sigma}_{\rm 2h} (< R) = \frac{2}{R^2} \int_0^{R} R' \bar{\Sigma}_{\rm 2h} (R') \operatorname{d}\!R' \mathrm{.}
\end{equation}

In practice, we integrate up to $50\si{Mpc}$ in $\bar{\Sigma}_{\rm 2h}(R)$, and verify that the function has converged at this point.
$\rho_{\rm c,0}$ and $\Omega_{\rm m,0}$ are respectively the critical density and the matter density at current time, and $\xi_{\rm gm}^{\rm 2h}$ is defined as
\begin{equation}
	\xi_{\rm gm}^{\rm 2h} (r) = b_{\rm h}(M) \zeta(r) \xi_{\rm m}(r)
\end{equation}
where $b_{\rm h}$ is the halo bias from \citet{seljak2004} and $\zeta(r)$ is the scale dependency of the bias as defined in equation B7 in \citet{tinker2005}:
\begin{equation}
	\zeta(r) = \frac{\left[ 1 + 1.17\xi_{\rm m}(r) \right]^{1.49} }{\left[ 1 + 0.69\xi_{\rm m}(r) \right]^{2.09} } \mathrm{.}
\end{equation}

$\xi_{\rm m}(r)$ is the non-linear matter correlation function, computed as in \citet{takahashi2012}, and using the linear matter correlation function from \citet{eisenstein1998}.

The two-halo halo term becomes predominant at scales of a few Mpc (see Fig \ref{fig:lensing_M1-2e12}), and has little influence  as we measure the lensing signal only up to $1.8h^{-1}\si{Mpc}$. We still include it to keep the model as generally applicable as possible.

The verification of the model predictions using numerical simulations are presented in appendix \ref{sec:simus}.

\section{Data}
\label{sec:data}

	\subsection{The source catalogues}

We use three shear catalogues to compute the weak lensing signal: CFHTLenS, CS82 and DES-SV, covering a total area of $393 \deg^2$. The total effective weighted source density is $n_{\rm eff} = \frac{1}{\Omega}\frac{\left(\sum w_i\right)^2}{\sum w_i^2} = 7.8$ galaxies/arcmin$^2$, with $\Omega$ the total effective area \citep{heymans2012}.

In the three catalogues, the estimation of photometric redshift, galaxy shape and stellar mass varies. To ensure that no catalogue gives biased results, we compare the results using each catalogue alone in appendix \ref{sec:cat_comp}.

		\subsubsection{CFHT Stripe 82}	

The CFHT Stripe 82 Survey \citep[CS82,][]{moraes2014} is a $i-$band imaging survey covering the $173 \deg^2$ of the SDSS Stripe 82 region, with a limiting magnitude of $i_{\rm AB} \sim 24.0$ and seeing between 0.4 and 0.8 arcsec. 

Each object has been attributed a \textsc{mask} flag indicating the quality of the photometry. Following \cite{erben2013} we use all objects with $\textsc{mask} \leq 1$. The remaining unmasked area is $129 \deg^2$.

We use photometric redshifts computed with BPZ \citep{benitez2000} from $ugriz$ SDSS photometry as described in \cite{bundy2015}.  The initial catalogue contains all the sources with a photometric redshift $z > 0$, but we test alternative cuts: we examine the balance between having more sources or having more confident redshifts by either keeping only sources with the parameter $\textsc{odds}>0.5$, or sources with $0.2 < z < 1.3$. The \textsc{odds} parameter is a measure of the peakiness of the redshift probability distribution around the best redshift, which means that objects with \textsc{odds} close to unity have a peaked and unimodal p(z), and thus have a more reliable redshift. After testing the different cuts we choose to apply $0.2 < z < 1.3$ for the rest of the work. We present the effect on our results of the different cuts in appendix \ref{sec:red_comp} \citep[see also][]{leauthaud2016}.
 
The galaxy shapes were measured by the CS82 collaboration as described in \citet{shan2017}, using the \texttt{lensfit} method \citep{miller2013}  following the procedure developed by the CFHTLenS collaboration \citep{ erben2009,erben2013}.
  The shear systematics and calibration verifications also follow CFHTLenS as described in \citet{heymans2012}. 
 Two calibration corrections need to be applied to the shear: one additive $c_2$  and one multiplicative $m$ (see section \ref{sec:lens_signal}). The pipeline also provides an inverse variance weight for each source, and objects with \textsc{weight} $=0$ are not used in the analysis. 
 We select galaxies using the object classification obtained with \texttt{lensfit}, which separates galaxies, stars and badly fit objects \citep[\textsc{fitclass} parameter, see details in][]{miller2013}.
 
In summary, we keep objects with $\textsc{mask} \leq 1$, $\textsc{weight} > 0$, $0.2 < z < 1.3$ and $\textsc{fitclass} = 0$.
The final source catalogue contains 3,791,129  sources with shear and photometric redshift measurement,  that is an effective weighted source density of $6.7$ galaxies/arcmin$^2$ (while $n_{\rm eff} = 12.3$ galaxies/arcmin$^2$ without any cuts). The completeness magnitude after the cuts is $i_{\rm AB} \sim 23.5$.

		\subsubsection{CFHTLenS}
			
The Canada-France-Hawaii Lensing Survey (CFHTLenS) is a $154\deg^2$ (146.5 after masking)  multi-band \textit{u*g'r'i'z'} survey based on the Wide component of the Canada-France-Hawaii Telescope Legacy Survey \citep{heymans2012}. 

 The photometric redshifts were  measured over the whole survey using the BPZ code with PSF-matched photometry \citep{hildebrandt2012}. We verify the effect on the lensing signal of the same source cuts as for the CS82 catalogue in appendix \ref{sec:red_comp}. Similarly to the CS82 catalogue, we finally use the cut $0.2 < z < 1.3$ for the rest of the work.
 
  The weak lensing data processing was performed with THELI \citep{erben2013}, and the shape measurement with \texttt{lensfit} \citep{miller2013}. The same calibration parameters as for the CS82 catalogue need to be applied to the shear.

In addition to the \textsc{fitclass} parameter, we cut our catalogue according to the \textsc{star\_flag} parameter to separate galaxies from stars. It classifies objects as galaxies or stars depending on their size, magnitude and colour information.

 Similarly to the CS82 catalogue, we keep objects with $\textsc{mask} \leq 1$, $\textsc{star\_flag} = 0$, $\textsc{fitclass} = 0$, $0.2 < z < 1.3$ and $\textsc{weight} > 0$. The final catalogue contains 5,615,617 sources on an effective  area after masking of $125 \deg^2$, which gives an effective weighted source density of 10.7 galaxies/arcmin$^2$ (and $n_{\rm eff} = 14.2$ galaxies/arcmin$^2$ without cuts).

		\subsubsection{DES Science Verification Data}
			
The Dark Energy Survey \citep[DES,][]{flaugher2005, flaugher2015, des2016} is an ongoing wide-field multi-band imaging survey, which will cover around $5000\deg^2$ in the grizY filters, to a depth of $i_{\rm AB} < 24$. 
In this analysis we use data taken during the Science Verification period (SVA1), which covers an area of $139 \deg^2$ after masking, with a depth comparable to the full survey\footnote{\url{https://des.ncsa.illinois.edu/releases/sva1}}. 

 We use the shear catalogue obtained using the \texttt{NGMix}\footnote{\url{https://github.com/esheldon/ngmix}} algorithm \citep{sheldon2014} and described in \cite{jarvis2015}, and photometric redshifts from \cite{bonnett2016}. After applying the cuts $\textsc{sva1\_flag} = 0$ (good galaxy) and $\textsc{ngmix\_flag} = 0$ (good shape measurement), the catalogue contains 3,446,533 galaxies, or a weighted density of $6.8$ galaxies/arcmin$^2$ ($n_{\rm eff} = 17.2$ galaxies/arcmin$^2$ without cuts). The completeness magnitude after the cuts is $i_{\rm AB} \sim 23$.

The multiplicative calibration factor, which needs to be applied to the shear as described in section \ref{sec:lens_signal}, is given in the \texttt{NGMix} catalogue as the sensitivity estimate \textsc{sens\_avg} and is related to the usual factor by $\textsc{sens\_avg} = 1 + m$.

The DES-SV data release contains a second catalogue with galaxy shape measurement, using an alternative algorithm \texttt{Im3shape}\footnote{\url{https://bitbucket.org/joezuntz/im3shape-git}}. We made some lensing measurements using both catalogues and found no significant difference, and thus chose to use the \texttt{NGMix} catalogue.

\begin{figure}
	\centering
	\includegraphics{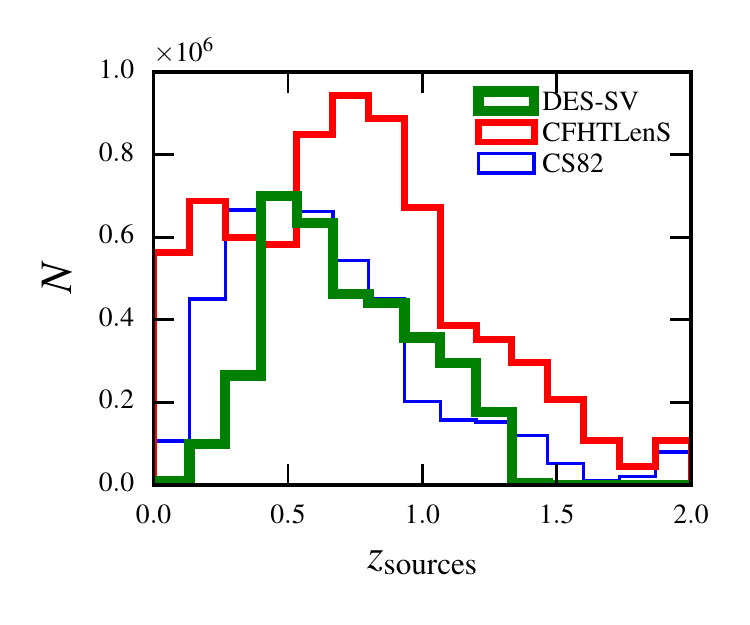}
	\caption{ Redshift  distribution of the source galaxies in the CS82, CFHTLenS and DES-SV surveys.
          }
	\label{fig:redshifts_sources}
\end{figure}

The redshift distributions of the sources in the three catalogues are presented in figure \ref{fig:redshifts_sources}.

	\subsection{Lens selection}
	\label{sec:lens_selection}
	
 Our sample consists of the overlap of the satellite galaxies from the redMaPPer catalogue with the CFHTLenS, CS82 and DES-SV fields. 
 
 The redMaPPer cluster catalogue is derived from photometric galaxy samples using the red-sequence Matched-filter Probabilistic Percolation cluster finder \citep{rykoff2014,rykoff2016}. For each galaxy the algorithm estimates the membership probability according to its color, position to the centre of the cluster and luminosity. To reduce contamination by line-of-sight galaxies, we select satellites with membership probability $P_{\rm mem} > 0.8$; as shown in \citet{zu2016b}, cutting at this value should eliminate most of the contamination. In our analysis we use the catalogues extracted from the 5-band $(ugriz)$  photometry of the SDSS Data Release 8 \citep[DR8,][]{sdss2011}, described in \citet{rykoff2014} and covering the CS82 and CFHTLenS fields, and from the $griz$ photometry of the Dark Energy Survey Science Verification Data  described in \citet{rykoff2016}.
  
\begin{figure}
	\centering
	\includegraphics{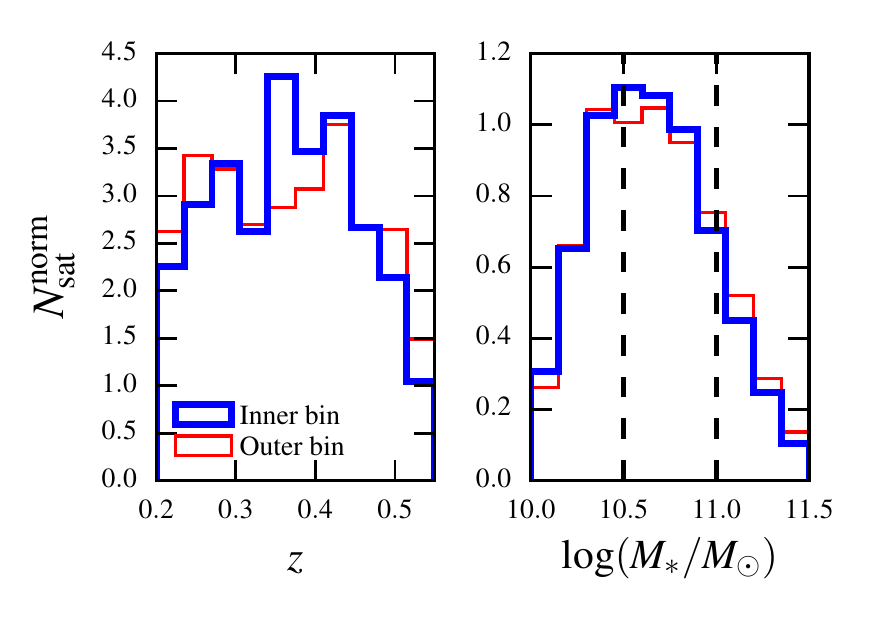}
	\caption{ Redshift (\textit{left panel}) and stellar mass (\textit{right panel}) distribution of the satellite galaxies in the combined catalogue, for the galaxies in the inner radial bin ($R_{\rm s} \in [0.1;0.55]h^{-1}\si{Mpc}$) in blue and the outer radial bin ($R_{\rm s} \in [0.55;1]h^{-1}\si{Mpc}$) in red. The distributions are normalized so their integrals sum to one, and the vertical lines in the right panel represent the limits of the stellar mass bins.
          }
	\label{fig:redshifts-Mstar}
\end{figure}

For the lenses we use the redMaPPer redshifts, as they are more robust than photometric redshifts estimated using other methods \citep[see][section 5]{bundy2015}. In order to have the same redshift range in the three samples, we keep only galaxies with $0.2 < z < 0.55$. The redshift distribution of the satellite galaxies in the two radial bins is shown in the left panel of figure \ref{fig:redshifts-Mstar}. In this redshift range, the redMaPPer cluster catalogue contains 289 clusters in the CFHTLenS footprint, 491 in CS82 and 349 in DES-SV. We note that the whole CFHTLenS field does not overlap with the redMaPPer cluster catalogue: there are around $30 \deg^2$ not covered, which gives an effective unmasked area of $\sim 115 \deg^2$ for CFHTLenS. The cluster density is $2.5\deg^{-2}$ in the CFHTLenS and DES-SV fields, and $3.8\deg^{-2}$ in the CS82 field. We suspect that this difference is due to the presence of a large structure in the CS82 field, as the cluster distribution appears to have a significant excess at $z\sim0.4$ compared to the other fields.
  
We select satellite galaxies according to their projected cluster-centric distance, in the two following bins: the satellites in the inner part of the clusters, with $R_{\rm s} \in [0.1; 0.55]h^{-1}\si{Mpc}$, and the satellites in the outer part of the clusters, with $R_{\rm s} \in [0.55; 1]h^{-1}\si{Mpc}$.  Indeed, the redMaPPer clusters used in the analysis have a richness between 20 and 180, which, according to the mass-richness relation from \cite{rykoff2012} corresponds to a virial mass between $10^{14}$ and $10^{15}$, which gives a virial radius in the range $0.6 - 1.4h^{-1}\si{Mpc}$. 

A possible alternative binning would be in fractions of $R_{\rm vir}$, ensuring that all the satellites are taken for each cluster. This choice would in addition avoid the selection effect discussed in section \ref{sec:cluster_mass}. However, we do not have precise estimations of the virial radius or mass for each cluster, and we should therefore use some mass-richness relation. These relations always show an important scatter \citep[see for example ][]{rozo2014, rozo2015a}, which would add noise to our bining.

We further split our lens sample in three stellar mass bins: the low mass sample with $\log(M_*/M_{\odot}) \in [10; 10.5]$, the intermediate mass sample with $\log(M_*/M_{\odot}) \in [10.5; 11]$ and the high mass sample with $\log(M_*/M_{\odot}) \in [11; 11.5]$. We have six lens samples, for each of which the number of lenses, mean stellar mass and mean redshift are shown in table \ref{tab:lenses}.
We describe in the next section how the stellar masses were obtained.

 	\subsection{Stellar mass}
	
 The CFHTLenS catalogue provides stellar masses for the galaxies, computed as described in \cite{velander2014},  by fitting spectral energy distribution (SED) templates, using the software \textsc{le Phare} \citep{arnouts1999, ilbert2006} following the method of \cite{ilbert2010}.

For CS82 and DES-SV we estimate the stellar masses using the software \textsc{le Phare} through the \textsc{gazpar}\footnote{\url{http://gazpar.lam.fr/index}} web service, and the available photometry for each survey:
\begin{itemize}
	\item  for the CS82 catalogue we use the  \textit{ugriz} SDSS Stripe 82 Coadd photometry and YJHK photometry from the UK Infrared Deep Sky Survey (UKIDSS) Large Area Survey;
	\item for DES-SV we use the \textit{griz} photometry.
\end{itemize}
	
For both catalogues the fitted SED templates are built from the stellar population synthesis (SPS) package from \citet{bruzual&charlot2003}, with a \citet{chabrier2003} initial mass function (IMF), and the star formation history is described as a decreasing exponential function  $e^{-t/\tau}$ with $0.1 \leq \tau \leq 30$Gyr. Dust extinction was applied following two laws, a starbust \citep{calzetti2000} and an intermediate slope ($\lambda^{0.9}$) law, and reddening excess $E(B-V)$ ranging from 0 to 0.7.

The stellar mass distribution  of the satellite galaxies in the two radial bins is shown in the right panel of figure \ref{fig:redshifts-Mstar}. The distributions for the satellites in the inner part of clusters and in the outer part are quite similar. Moreover, assuming that the stellar masses of galaxies do not vary much on average during infall \citep[due to the star formation quenching and the relative insensitivity of the stellar component to stripping, see][]{smith2016}, we consider that any measured mass difference between the satellites in the two radial bins is due to the evolution of the subhaloes during accretion.
We study the validity of the assumption that the stellar mass remains constant during accretion in section \ref{sec:star_evol}.

 	\subsection{Computation of the lensing signal}
	\label{sec:lens_signal}
	
We measure $\Delta\Sigma$ by stacking lens-source pairs in 13 radial bins from 0 to $1.8h^{-1}\si{Mpc}$. For each sample of lenses, $\Delta\Sigma(R)$ is estimated using
\begin{equation}
\Delta\Sigma(R) = \frac{\sum_{\rm ls}w_{\rm ls}\gamma_{\rm t}^{\rm ls}\Sigma_{\rm crit}(z_{\rm l}, z_{\rm s})}{\sum_{\rm ls}w_{\rm ls}} \mathrm{,}
\end{equation}
where $\gamma_{\rm t}^{\rm ls}$ is the tangential shear, $w_{\rm ls} = w_{\rm s}/\left(\Sigma_{\rm crit}\left(z_{\rm l}, z_{\rm s}\right)\right)^{2}$, and $w_{\rm s}$ is an inverse variance weight factor associated to each source galaxy and introduced to account for shape measurement error and intrinsic scatter in galaxy ellipticity \citep{heymans2012}. The sum is calculated over all the lens-source pairs.

The multiplicative calibration factor $m$ needs to be taken into account in a statistical way \citep{miller2013}, we apply it to the mean shear measurement using the correction proposed in \citet{velander2014} and \citet{hudson2015}:
\begin{equation}
1 + K(z_{\rm l}) = \frac{\sum_{\rm ls} w_{\rm ls}(1+m)}{\sum_{\rm ls} w_{\rm ls}}\mathrm{,}
\end{equation}
which gives a calibrated lensing signal:
\begin{equation}
\Delta\Sigma^{\rm cal}(R) = \frac{\Delta\Sigma(R)}{1 + K(z_{\rm l})} =  \frac{\sum_{\rm ls}w_{\rm ls}\gamma_{\rm t}^{\rm ls}\Sigma_{\rm crit}} {\sum_{\rm ls} w_{\rm ls}(1+m)} \mathrm{.}
\end{equation}

We only use lens-source pairs with $z_{\rm source} > z_{\rm lens} + z_{\rm lens}^{\rm err} + z_{\rm source}^{\rm err}$ to ensure that no sources are at lower redshift than the lens. 
The error bars on the lensing signal are obtained with a block bootstrap on the data: the field is divided in blocks, and the lensing signal is measured on the resampled blocks to estimate the variance of the measurement. 

We compute the lensing signal for each lens sample, using a modified version of the \texttt{athena}\footnote{\url{http://www.cosmostat.org/software/athena/}} software, a 2d-tree code estimating second-order correlation functions from input galaxy catalogues.

\section{Results}
\label{sec:results}

	\subsection{Fitting procedure}
	
We fit the lensing signal with the model described in section \ref{sec:model}, which has two free parameters: the mean satellite mass $M_{\rm sat}$ and the mean host mass $M_{\rm host}$. 
We obtain the best-fit parameters by maximizing the likelihood  $\mathcal{L}$, and to obtain the intervals of confidence we use a Markov chain Monte Carlo (MCMC) method using \textsc{emcee} \citep{foreman-mackey2013} which is a Python implementation of an affine invariant MCMC ensemble sampler.

The likelihood $\mathcal{L}$ is expressed as:
\begin{equation}
\mathcal{L} = \frac{1}{\sqrt{2\pi |C|}}\exp(-\frac{1}{2}\sum_{i=1}^{13}\frac{(x^{\rm obs}_{i} - x^{\rm mod}_{i})^2}{\sigma_i^2})
\end{equation}
where  $x^{\rm obs}_i$ are the measurements and $x^{\rm mod}_i$ the model predictions in the 13 radial bins. The $\sigma_i$ are the diagonal terms of the covariance matrix $C$ computed with the block bootstrap. We use only the diagonal as the full matrix is quite noisy. 
The average amplitude of a non-diagonal term of the covariance matrix is around $10\%$ of a diagonal term, which is consistent with the noise in matrices computed with bootstrap as shown in \citet{viola2015}.  The use of only diagonal terms might result in overestimating the quality of the fit, which could underestimate the size of the errorbars on the mass measurements. 
We assume flat and broad priors: for the satellite mass we choose $\log(M_{\rm sat}/h^{-1}M_{\odot}) \in [9.5;13.5]$ and for the cluster mass $\log(M_{\rm host}/h^{-1}M_{\odot}) \in [13.5;16]$.

	\subsection{Dependance of the SHMR on the projected cluster-centric distance}
	\label{sec:shmr}

\begin{figure*}
	\centering
	\includegraphics{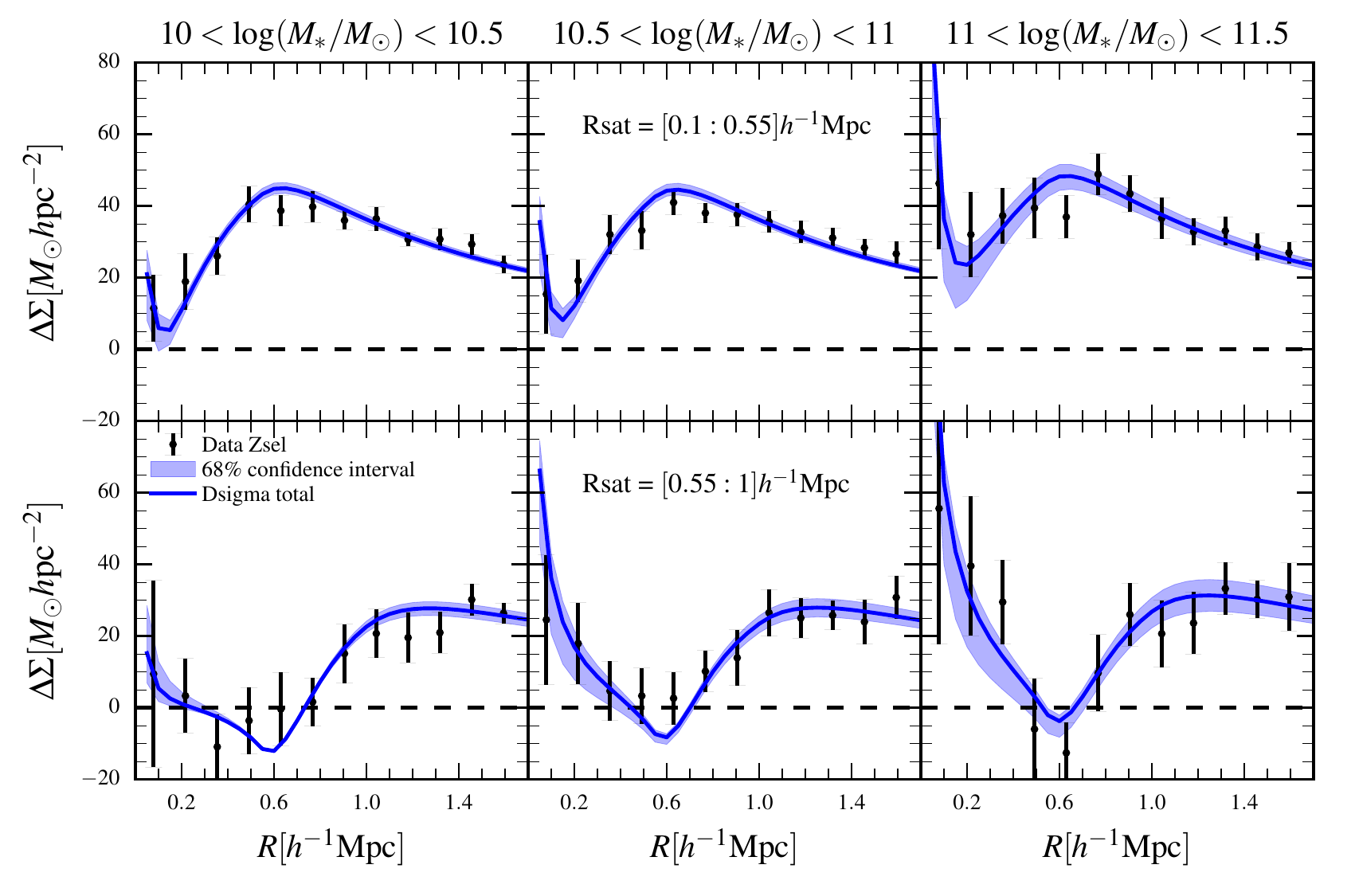}
	\caption{ Lensing by the redMaPPer satellite galaxies. The top line shows the lensing signal for the satellites in the inner part of the cluster, with $R_{\rm s} \in [0.1; 0.55]h^{-1}\si{Mpc}$, and the bottom one for the satellites in the outer part, with $R_{\rm s} \in [0.55; 1]h^{-1}\si{Mpc}$. The left column is for satellites with $\log(M_*/M_{\odot}) \in [10; 10.5]$, the middle column for $\log(M_*/M_{\odot}) \in [10.5; 11]$ and the right one for $\log(M_*/M_{\odot}) \in [11; 11.5]$.  The black dots are the data points with bootstrap errors, the blue curve is the best-fit model, and the blue surface shows the 68\% confidence interval. 
          }
	\label{fig:all_lensing}
\end{figure*}

We plot the lensing signal for each of the lens samples in figure \ref{fig:all_lensing}, with columns from left to right showing the bins in log stellar mass [10;10.5], [10.5;11], [11;11.5]. The top line shows the signal for satellites in the inner parts of clusters ($R_{\rm s} \in [0.1;0.55]h^{-1}\si{Mpc}$), and the bottom line for satellites in the outer part ($R_{\rm s} \in [0.55;1]h^{-1}\si{Mpc}$).  In addition, for each sample we plot the best-fit model: the blue continuous line is the full model, and the blue shaded area is the $68\%$ credible interval.

\begin{figure*}
\centering

\begin{tabular}{ccc}
\subfloat[$10 < \log M_* < 10.5$, $0.1 < R_{\rm s} < 0.55$]{\includegraphics[height=5.5cm]{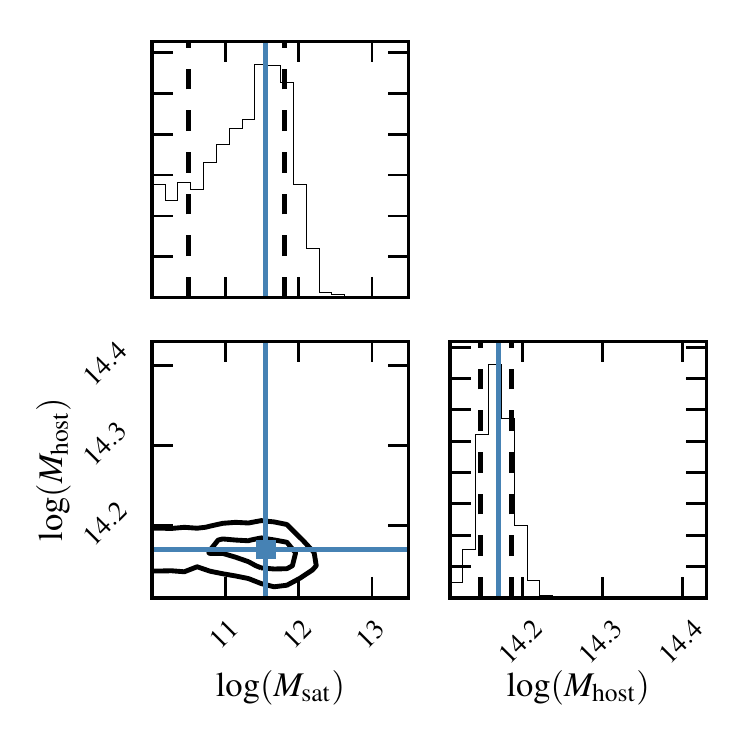}}   & \subfloat[$10.5 < \log M_* < 11$, $0.1 < R_{\rm s} < 0.55$]{\includegraphics[height=5.5cm]{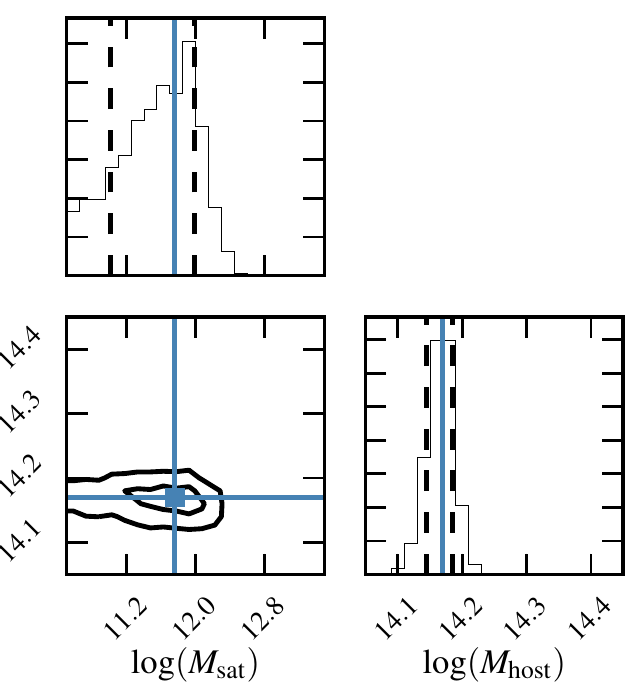}} & \subfloat[$11 < \log M_* < 11.5$, $0.1 < R_{\rm s} < 0.55$]{\includegraphics[height=5.5cm]{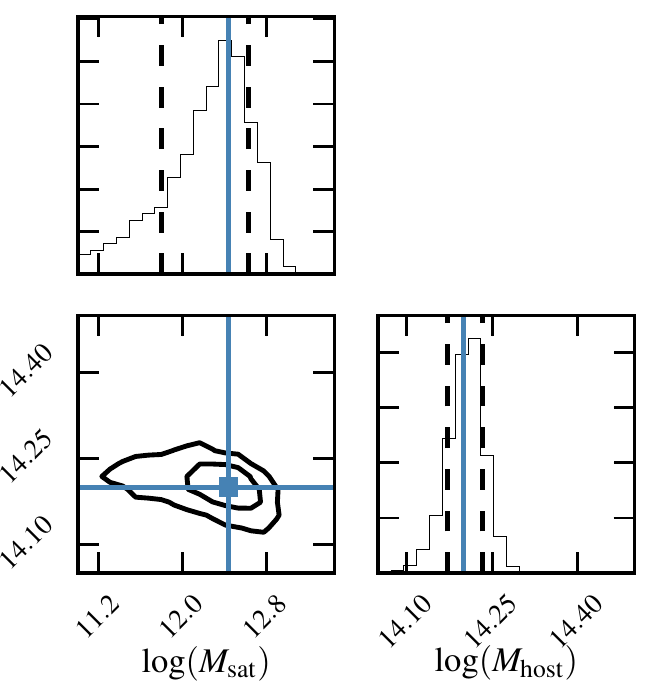}}\\
\subfloat[$10 < \log M_* < 10.5$, $0.55 < R_{\rm s} < 1$]{\includegraphics[height=5.5cm]{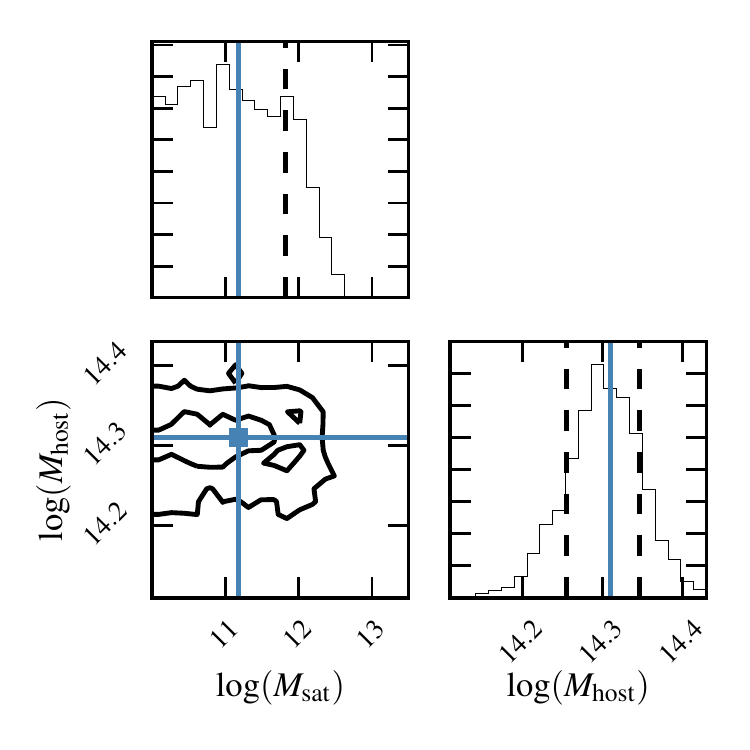}}  & \subfloat[$10.5 < \log M_* < 11$, $0.55 < R_{\rm s} < 1$]{\includegraphics[height=5.5cm]{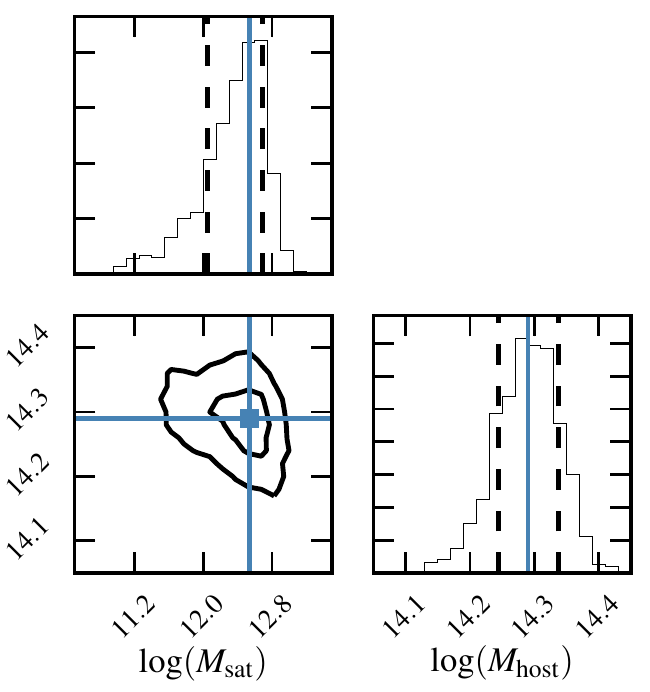}}  & \subfloat[$11 < \log M_* < 11.5$, $0.55 < R_{\rm s} < 1$]{\includegraphics[height=5.5cm]{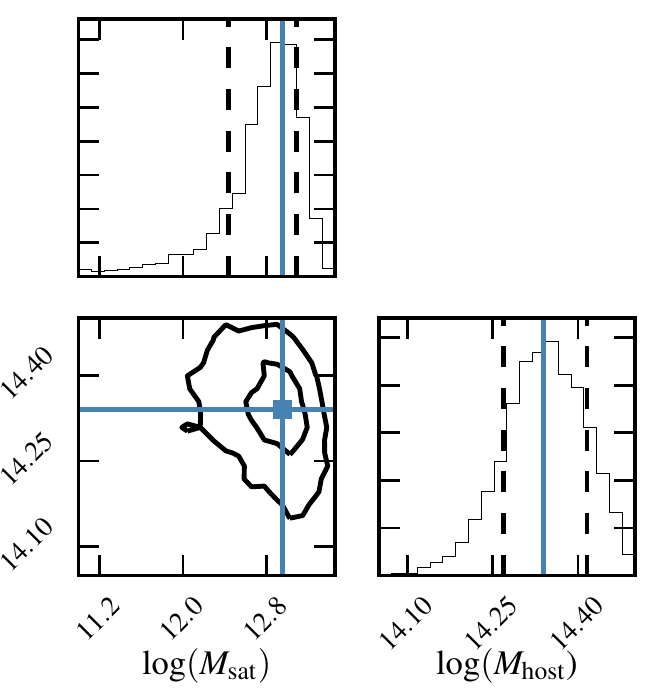}}\\ 
\end{tabular}


\caption{ Joint 2-dimensional and marginalized 1-dimensional posterior probability distributions for our two parameters $M_{\rm sat}$ and $M_{\rm host}$ for each of the six samples: the top line presents the bins with $R_{\rm s} \in [0.1;0.55]h^{-1}\si{Mpc}$, and the bottom line $R_{\rm s} \in [0.55;1]h^{-1}\si{Mpc}$. The columns, from left to right represent the stellar mass bins $\log(M_{*}/M_{\odot}) \in [10-10.5], [10.5-11], [11-11.5]$. In the 1d distributions, the blue lines are the maximum likelihood parameters, and the black dashed lines are the 68\% credible intervals. In the 2d distributions, the blue crosses are the maximum likelihood solutions, and the black contours are the 68\% and 95\% joint credible regions. Masses are in units of $h^{-1}M_{\odot}$ and distances in $h^{-1}\si{Mpc}$.}\label{fig:corners}
\end{figure*}

The maximum likelihood results with the $68\%$ credible intervals are presented in table \ref{tab:lenses}, with the average stellar mass and redshift for each lens sample.
In addition, we present the joint 2-dimensional and marginalized 1-dimensional posterior probability distributions for the two parameters $M_{\rm sat}$ and $M_{\rm host}$ for each of the six samples in figure \ref{fig:corners}.

\begin{figure}
	\centering
	\includegraphics{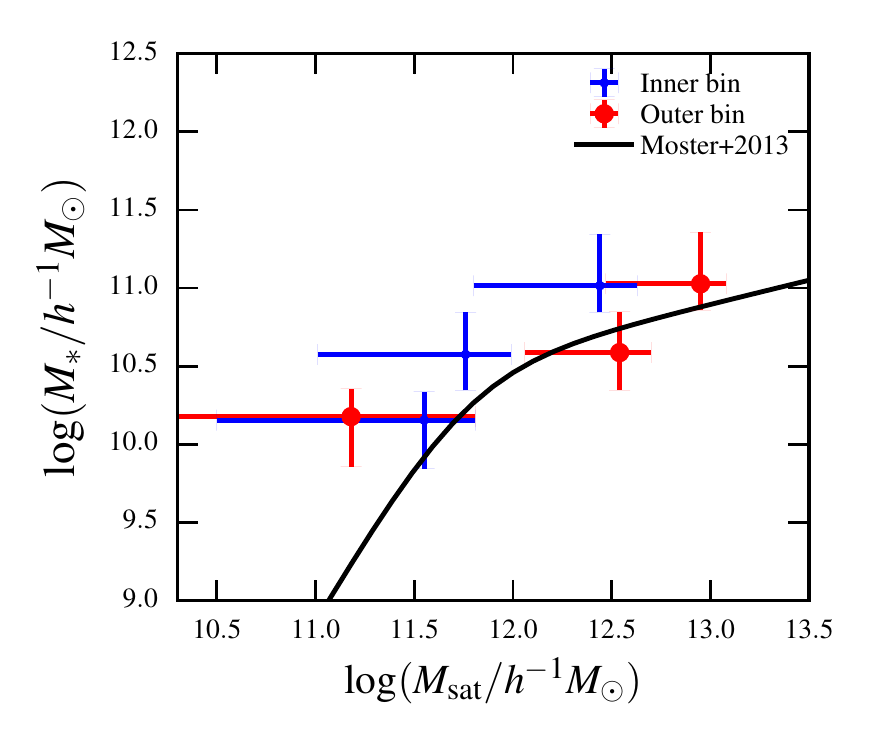}
	\caption{Stellar-to-halo mass relation measured for the satellite galaxies in our sample, in blue for the galaxies in the inner part of clusters ($R_{\rm s}\in[0.1;0.55]h^{-1}\si{Mpc}$) and in red for the galaxies in the outer part of clusters ($R_{\rm s}\in[0.55;1.]h^{-1}\si{Mpc}$). The black line is the SHMR for field/central galaxies at $z=0.35$, computed from simulations in \citet{moster2013}.
          }
	\label{fig:shmrelation}
\end{figure}

For the low mass satellites, the signal-to-noise is too low to obtain strong constraints.  For the intermediate and high mass samples, we find that at a given stellar mass, the dark matter halo mass  is shifted toward lower mass for satellites in the inner part of clusters, compared to the satellites in the outer part.
 A summary of the measured satellite dark matter mass as a function of the mean stellar mass in each sample is shown in figure \ref{fig:shmrelation}.
	
\begin{table*}
\centering
\begin{tabular}{c c c c c c c}
$M_*$ range [$M_{\odot}$]	&	$R_s [h^{-1}\si{Mpc}]$	&	$N_{\rm lenses}$	&	$<\log{M_{*}/M_{\odot}}>$	&	$<z_{\rm l}>$	&	$\log(M_{\rm sat}/h^{-1}M_{\odot})$	&	$\log(M_{\rm host}/h^{-1}M_{\odot})$	\\
\hline
\multirow{2}{*}{10-10.5}	&	0.1-0.55			&	48891		&	10.31				&	0.35		&	$11.55^{+0.26}_{-1.05}$		&	$14.17^{+0.02}_{-0.02}$			\\
					&	0.55-1			&	1292			&	10.32				&	0.37		&	$11.18^{+0.63}_{-1.26}$		&	$14.31^{+0.04}_{-0.05}$			\\
\hline
\multirow{2}{*}{10.5-11}	&	0.1-0.55			&	6935			&	10.73				&	0.37		&	$11.76^{+0.23}_{-0.75}$		&	$14.17^{+0.02}_{-0.02}$			\\
					&	0.55-1			&	1836			&	10.73				&	0.36		&	$12.54^{+0.16}_{-0.48}$		&	$14.29^{+0.05}_{-0.05}$			\\
\hline
\multirow{2}{*}{11-11.5}	&	0.1-0.55			&	2126			&	11.17				&	0.38		&	$12.44^{+0.19}_{-0.64}$		&	$14.20^{+0.03}_{-0.03}$			\\
					&	0.55-1			&	677			&	11.17				&	0.36		&	$12.95^{+0.13}_{-0.48}$		&	$14.34 ^{+0.08}_{-0.07}$			\\
\hline
\end{tabular}
\caption{Description of the lens samples and best-fit parameters.}
\label{tab:lenses}
\end{table*}

 We expect the satellites in the outer parts of the clusters to be similar to field/central galaxies. As the distance to the centre of the cluster correlates with the accretion redshift, most of them have only started recently their accretion process, and have not yet been submitted to strong tidal forces from the cluster for a long time. On the contrary, the satellites close to centre of the cluster have undergone the influence of the cluster much more strongly and for a longer time, and have thus been stripped of a significant part of their dark matter. This hypothesis is in agreement with our measurements. 
 We plot on figure \ref{fig:shmrelation} the stellar-to-halo mass relation for field/central galaxies measured from N-body simulations in \citet{moster2013}. Their relation is indeed consistent with the one we measure for satellites in the outskirts of clusters. In addition, the shift of the SHMR towards lower subhalo mass is consistent with the results obtained in \citet{rodriguez2012, rodriguez2013} using extended abundance matching technique.
  
Assuming that our measurements do reflect the dark matter stripping scenario, we quantify the effect for the two higher mass bins by computing stripping factors which represent the amount of stripped dark matter, defined as 
\begin{equation}
	\tau_{\rm strip} = 1 - \frac{M_{\rm sat}^{\rm inner}}{M_{\rm sat}^{\rm outer}} \mathrm{.}
\end{equation}
For the intermediate mass bin, we obtain $\tau_{\rm strip}^{\rm inter} = 0.83^{+0.15}_{-0.69}$, and for the high mass $\tau_{\rm strip}^{\rm high} = 0.69^{+0.26}_{-1.14}$. Our error bars do not allow us to have strong constrains, but we still compute a theoretical value for the stripping factor as a comparison. Using the equation from \citet{gao2004} for the infall mass, the stripping rate can be expressed as:
\begin{equation}
	\tau_{\rm strip}(R_{\rm s}) = 1 - \frac{M_{\rm sub}(R_{\rm s})}{M_{\rm infall}} = 1 - 0.65\left(\frac{R_{\rm s}}{R_{\rm vir}}\right)^{2/3}\mathrm{.}
\end{equation}
We take for the distance cluster centre-satellite the median value in the inner bin sample $R_{\rm s} = 0.28h^{-1}\si{Mpc}$ and for the host virial radius the value corresponding to the median between our best-fit host masses $R_{\rm vir} = 1.16h^{-1}\si{Mpc}$, and obtain a stripping rate $\tau_{\rm strip} = 0.75$. 

We compute the relative likelihood of obtaining a value of $\tau_{\rm strip}$ equal to 0 (no stripping) or 0.75 (theoretical value for stripping) using our posterior probability distributions as 
\begin{multline}
\mathcal{L}(\tau ) = \mathcal{L}\left(M_{\rm sat}^{\rm inner} = (1-\tau)M_{\rm sat}^{\rm outer}\right) \\
		      = \int \mathcal{L}\left(M_{\rm inner} = (1-\tau)M'\right) \times \mathcal{L}\left(M_{\rm outer} = M'\right)dM' 
\end{multline}
assuming the independence of our samples. We obtain for the high mass sample $P(\tau= 0) = 0.43$ and $P(\tau=0.75) = 0.84$, and for the intermediate mass sample $P(\tau= 0) = 0.28$ and $P(\tau=0.75) = 0.59$, which shows that our results clearly favour the stripping scenario over the no stripping one. In addition we perform a Kolmogorov-Smirnov two-sample test to verify if the posterior probability distributions for $M_{\mathrm{sat}}^{\mathrm{inner}}$ and $M_{\mathrm{sat}}^{\mathrm{outer}}$ are statistically different, and reject the null hypothesis (no stripping) since the p-value is below $1\%$.

 

\begin{figure}
	\centering
	\includegraphics{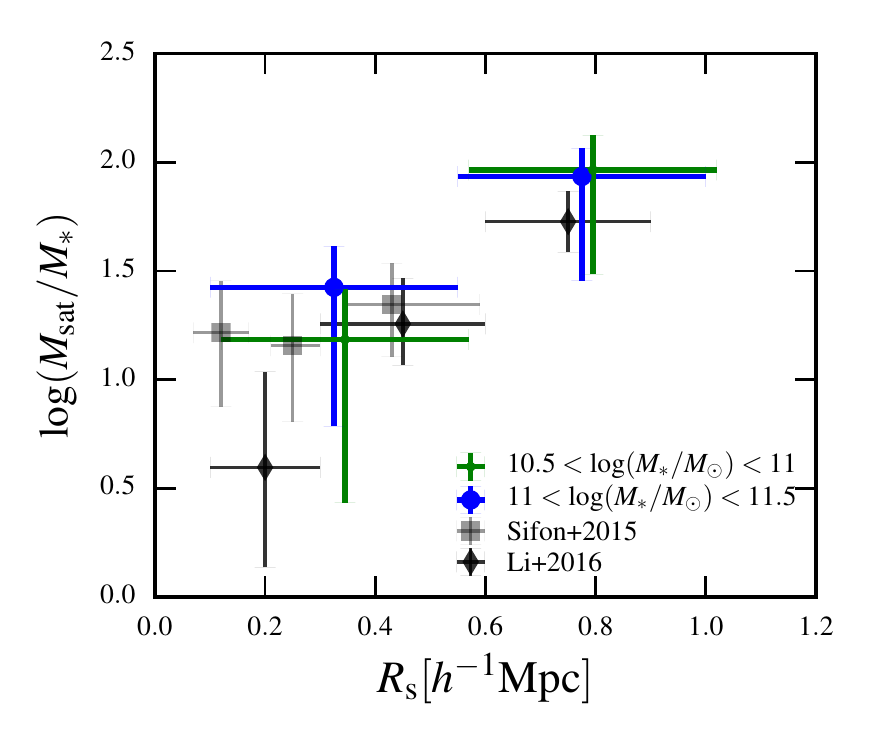}
	\caption{Subhalo to stellar mass ratio as a function of the cluster-centric distance, calculated for the satellites in the intermediate (green) and high (blue) stellar mass bins. We do not plot the low stellar mass bin as it is not very constrained. We plot as a comparison the results from \citet{li2015} and \citet{sifon2015}.
          }
	\label{fig:shmratio}
\end{figure}

Finally, we compute the ratio subhalo mass over stellar mass for the subhaloes in the intermediate and high stellar mass bins, and plot it in figure \ref{fig:shmratio}. We also plot the results from \citet{sifon2015} and \citet{li2015} and find them to be broadly consistent with our results.

	\subsection{Mass of the redMaPPer clusters}
	\label{sec:cluster_mass}

We now look at our measurement of the mass of the host redMaPPer clusters. In each of the cluster-centric bins, we find that the best-fit host halo mass has a consistent value across the different stellar mass bins. However, the average host mass is lower for satellites in the inner part of clusters ($\log M_{\rm host} \sim 14.20$) than for satellites in the outskirts ($\log M_{\rm host} \sim 14.30$).
This is due to a selection effect: larger host clusters have more satellites at a distance ranging between 0.55 and 1$h^{-1}\si{Mpc}$, and are thus more represented in the outer part satellite bin, pushing the mean host halo mass towards higher values.

We verify this assumption by computing the mean host richness in each satellite bin. For galaxies in the inner part of clusters, we find that the mean richness of their host varies between 43 and 45 (depending on the stellar mass bin), while for galaxies in the outer part the mean richness is between 58 and 63.

In addition, we compute from the richness the halo mass using the mass-richness relation from \citet{rykoff2012}: we find that for the satellites in the centre the mean host halo mass in each of the different samples computed from the mean richness is in the range $\log(M_{\rm host}/h^{-1}M_{\odot}) \in [14.21;14.22]$, and for the satellites in the outskirts the mean host mass is in $\log(M_{\rm host}/h^{-1}M_{\odot}) \in [14.31;14.34]$, which agrees very well with our mass measurements given in table \ref{tab:lenses}.

\section{Discussion}
\label{sec:discussion}

	\subsection{Stellar mass evolution during infall}
	\label{sec:star_evol}
	
	In this study, our observations suggest that for a given stellar mass, satellite galaxies have a more massive dark matter halo if they are in the outskirts of their host cluster than if they are close to its centre. We explain these observations with the following scenario: during its accretion to a cluster, a galaxy has its star formation quenched while its dark matter halo is stripped by the tidal forces of the host. In this section, we consider whether some alternative infall processes could be considered, such as evolution of the satellite stellar mass.
	
A first possibility is that if the stellar component of the galaxy gets stripped along the dark matter halo, the stellar mass and the dark matter mass would decrease during infall. 
However, as tidal forces are expected to remove matter from a galaxy starting from the outside towards the centre, stellar matter is expected to remain undisturbed much longer than dark matter. Using hydrodynamical simulations, \citet{smith2016} studied how the stellar component of a galaxy is affected by stripping compared to the dark matter halo. Using their equation 2, we compute the fraction of remaining stellar matter after stripping, assuming that dark matter has been stripped by a factor of $\tau_{\rm strip}$. Using the best-fit values obtained in section \ref{sec:shmr}, we find that for the high mass sample $99\%$ of the stellar component remains, and $91\%$ for the intermediate mass sample. The possibly removed amount of stellar matter is thus very small and would not change our conclusions. Even if the equation from \citet{smith2016} underestimates the amount of stellar stripping, considering that stars are partly stripped during accretion only strengthens our conclusions: if a satellite galaxy has been stripped of $dM_*$ stars during infall, its progenitor was a galaxy of stellar mass $M_* + dM_*$ which would have a more massive dark matter halo than a progenitor of stellar mass $dM_*$ \citep[assuming that the relation between stellar mass and dark matter mass is monotonic, as in][]{moster2013}. 
	
We now focus on the more interesting topic of star formation during accretion. Indeed, what we explain as a stripping of the dark matter at constant stellar mass might also be an increase of the stellar mass at fixed (or less shifted) dark matter mass. We verify here how this scenario would fit with our observations, by estimating how much stars could have formed during the infall process for our sample of galaxies in the inner part of clusters. First, using the equation from \citet{giocoli2008} we estimate the time \textit{since} accretion $t_{\rm inf}$ and the redshift \textit{at} accretion $z_{\rm inf}$:
\begin{equation}
	t_{\rm inf} = \ln\left[0.65\left(\frac{R_{\rm s}}{R_{\rm vir}}\right)^{2/3}\right]t_{\rm dyn}(z)\mathrm{,}
\end{equation}
where \citep{vandenbosch2005b}
\begin{equation}
	t_{\rm dyn}(z) = 2\left(\frac{\Delta_{\rm vir}(z)}{\Delta_{\rm vir}(0)}\right)^{-0.5}\frac{H(0)}{H(z)}\mathrm{,}
\end{equation}
where $\Delta_{\rm vir}(z)$ is the virial overdensity at redshift $z$. Using the same values for $R_{\rm s}$ and $R_{\rm vir}$ as for the stripping factor in section \ref{sec:shmr}, we find $t_{\rm inf} = 2.03\si{Gyr}$ and $z_{\rm inf} = 0.62$.

We can then obtain the amount of star formation that occurred during the infall time, assuming that there is no quenching due to the cluster environment at all. We use star formation rates computed in \citet{buat2008}, that is, for the three stellar mass bins, from low to high: $SFR(M_*^1, z_{\rm inf}) \sim 5 \si{M}_{\odot}\si{/yr}$,  $SFR(M_*^2, z_{\rm inf}) \sim 11\si{M}_{\odot}\si{/yr}$ and  $SFR(M_*^3, z_{\rm inf}) \sim 23 \si{M}_{\odot}\si{/yr}$. 

\begin{figure}
	\centering
	\includegraphics{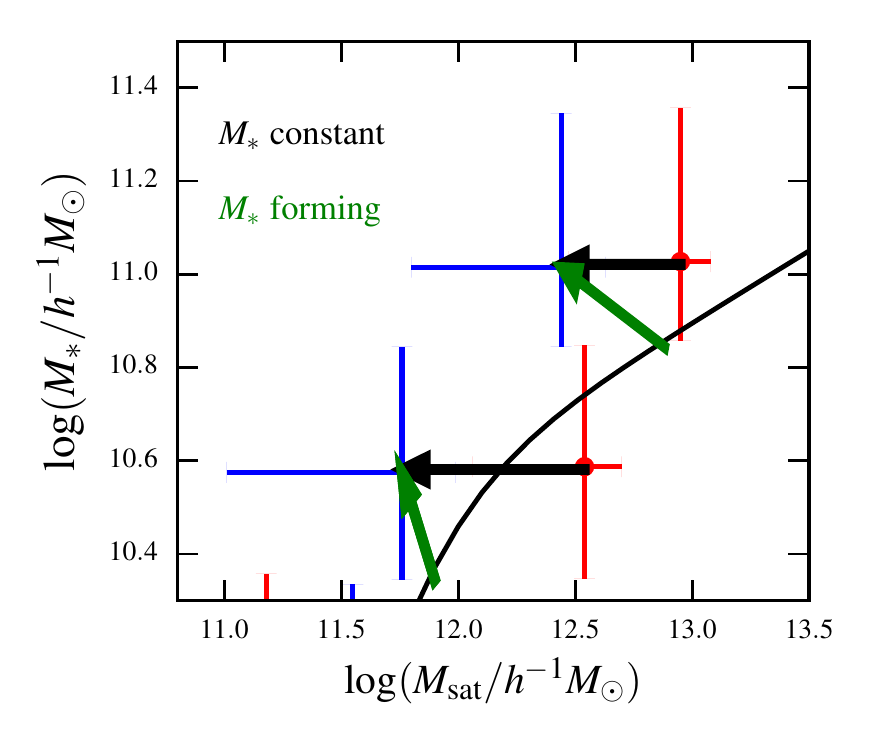}
	\caption{Same as figure \ref{fig:shmrelation} zoomed on the two higher stellar mass bins. We show two possible scenario to explain our observations: the black arrows would represent a stripping of the dark matter at fixed stellar mass, while the green arrows represent a coevolution of the stellar and dark matter, with star formation rates from \citet{buat2008}.
          }
	\label{fig:shmr_alt}
\end{figure}

We can then compute the stellar mass at infall for the three stellar mass bins: 
\begin{equation}
	M_{*\rm ,inf} = M_{*} - t_{\rm inf}*SFR(z_{\rm inf})\mathrm{,}
\end{equation}
which gives $\log(M_{*\rm ,inf}^1/M_{\odot}) = 10.00$, $\log(M_{*\rm ,inf}^2/M_{\odot}) = 10.50$ and $\log(M_{*\rm ,inf}^3/M_{\odot}) = 11.00$. Using the relation from \citet{moster2013}, we infer the corresponding dark matter halo mass before accretion, and show in figure \ref{fig:shmr_alt} the corresponding evolution scenario compared to the scenario with constant stellar mass. In the case where we consider an evolution in the stellar mass (green arrows), the shift in dark matter mass is less important than for the case with fixed stellar mass (black arrows), but still consistent with some stripping of the dark matter halo. 

It is important to note that we use here star formation rates that correspond to star-forming galaxies \citep{buat2008} and that many studies show that cluster galaxies are at least partly quenched in this dense environment. The two scenarios we present are therefore only the two extreme limits (with no star-formation and with no quenching), and a more detailed study of the coevolution of stellar and dark matter should be carried out.

	\subsection{Systematic error}

We perform different tests to verify the robustness of our results. The analyses are described in appendix \ref{sec:num_sim} and \ref{sec:syst_err}, and we summarize here the conclusions.

To verify if no systematic bias is introduced in the lensing profiles by the differences in the source catalogues (shape measurements, redshift estimations, etc.) we perform the same measurement using each catalogue individually.  The subhalo masses agree within one sigma between the different lensing catalogues.

We then examine the influence of the redshift quality in the CS82 and CFHTLenS catalogues, by testing different quality cuts and comparing the results. We chose for our fiducial source catalogue all galaxies with $0.2 < z < 1.3$.

Finally, using the BigMultiDark simulations  \citep[described in appendix \ref{sec:simus}]{giocoli2016, klypin2016}, we test the influence of line-of-sight projections on our measurements. In the redMaPPer cluster catalogue we cut satellites according to their cluster membership probability to improve the purity of the satellite catalogue, but some contamination by non-member galaxies is still possible. We quantify the effect of this on the lensing signal using simulations, and find that contamination can decrease the amplitude of the lensing profile by $\sim 1\%$.

\section{Conclusion}
\label{sec:conclusion}
 
 Many numerical simulations and observations suggest that dark matter haloes of satellite galaxies are subjected to the tidal stripping induced by the gravitational potential of their host cluster. In this work, we study this effect by measuring the mass of satellite galaxies using galaxy-galaxy lensing: we compare the mass of satellites located in the outer part of clusters (which have just started their accretion process and are therefore less perturbed by the cluster), with satellites in the inner parts of clusters (which, in average, have been subject to the influence of the cluster for a longer time). We further divide our satellites in stellar mass bins, as the stellar mass of a galaxy appears to be a good tracer of the mass of the galaxy before infall.
 
 We find good agreement with \citet{moster2013} for the mass of satellite galaxies in the outer radial bin and a suggested shift to smaller subhalo masses in the inner radial bin, in agreement with the dark matter stripping scenario.  For the intermediate and high stellar mass bins, we find stripping factors of $\tau^{\rm inter}_{\rm strip} = 83^{+15}_{-69} \%$ and $\tau^{\rm high}_{\rm strip} = 69^{+26}_{-114} \%$. 
 Using  the posterior probability distributions, we find that the theoretical stripping scenario $\tau_{\rm strip} = 0.75$ is favoured with respect to the no stripping scenario $\tau_{\rm strip} = 0$. We find for the intermediate mass sample $P(\tau_{\rm strip} = 0.75) = 0.59$ and $P(\tau_{\rm strip} = 0) = 0.28$, and for the high mass sample  $P(\tau_{\rm strip} = 0.75) = 0.84$ and $P(\tau_{\rm strip} = 0) = 0.43$.
  
 While our estimations of the subhalo masses appear to be consistent with a stripping of the dark matter by the gravitational potential of the host clusters, some other effects can have an influence on the results. We consider two possible evolutions in stellar mass, stripping or star-formation, and find that while stellar stripping can be neglected in the process, stellar formation could partly explain our observations. To disentangle the contributions of dark matter stripping and star formation, a study of the coevolution of dark and stellar matter during accretion should be carried out.

To improve the measurements of the subhalo mass, different solutions can be considered. Future lensing surveys will cover thousands of square degrees instead of hundreds, decreasing statistical errors. In addition, spectroscopically confirmed clusters can be studied, to remove contamination and have more precise redshifts. The improvement of the lensing measurement of subhaloes will allow not only to constrain their mass evolution more precisely, but also to study their density profiles and measure the tidal truncation radius.

 \section*{Acknowledgements}
 Based on observations obtained with MegaPrime/MegaCam, a joint project of CFHT and CEA/DAPNIA, at the Canada-France-Hawaii Telescope (CFHT), which is operated by the National Research Council (NRC) of Canada, the Institut National des Science de l'Univers of the Centre National de la Recherche Scientifique (CNRS) of France, and the University of Hawaii. The Brazilian partnership on CFHT is managed by the Laborat\'orio Nacional de Astrof\'isica (LNA). We thank the support of the Laborat\'orio Interinstitucional de e-Astronomia (LIneA). We thank the CFHTLenS team. This work was granted access to the HPC resources of Aix-Marseille Universit\'{e} financed by the project Equip@Meso (ANR-10-EQPX-29-01) of the program "Investissements d'Avenir" supervised by the Agence Nationale pour la Recherche. This work is based in part on data products produced by GAZPAR located at the Laboratoire d'Astrophysique de Marseille. M. L thanks CNRS for financial support. MM is partially supported by CNPq and FAPERJ. Fora Temer. This work is supported by the Deutsche Forschungsgemeinschaft in the framework of the TR33 'The Dark Universe'. Carlo Giocoli thanks CNES and acknowledges support from the Italian Ministry for Education, University and Research (MIUR) through the SIR individual grant SIMCODE, project number RBSI14P4IH. HH was supported by an Emmy Noether grant of the Deutsche Forschungsgemeinschaft (Grant no.: Hi 1495/2-1).

 \appendix
 
 \section{Numerical simulations}
 \label{sec:num_sim}
 
 	\subsection{Description of the simulations}
	 \label{sec:simus}

We use the BigMultiDark Planck cosmological simulation \citep[BigMDPL,][\url{www.multidark.org}]{klypin2016} which contains $3840^3$ particles in a box of comoving side $2.5h^{-1}Gpc$. Haloes and subhaloes are identified from density peaks using the Bound Density Maximum (BDM) halo finder \citep{klypin&holtzman1997, riebe2013}.
The construction of the light-cones is described in \citet{giocoli2016}: the particles from 24 snapshots of the simulation were projected onto 24 lens planes distributed along the line-of-sight up to redshift $z = 2.3$. To simulate lensing, they trace light-rays through the successive lens planes of the light-cones using the ray-tracing code \textsc{Glamer} \citep{metcalf2014, petkova2014}.

The simulation mass resolution (ie mass of the smallest particle) is $2.35\times10^{10}h^{-1}M_{\odot}$. The effect of this limited mass resolution has an impact on the lensing measurements, and induce a damping of the signal at small scales.

For our measurements we use 15 lightcone realizations. Each realization corresponds to a rectangular field covering $55 \deg^2$.

	\subsection{Model verification}
 
 \begin{figure}
	\centering
	\includegraphics{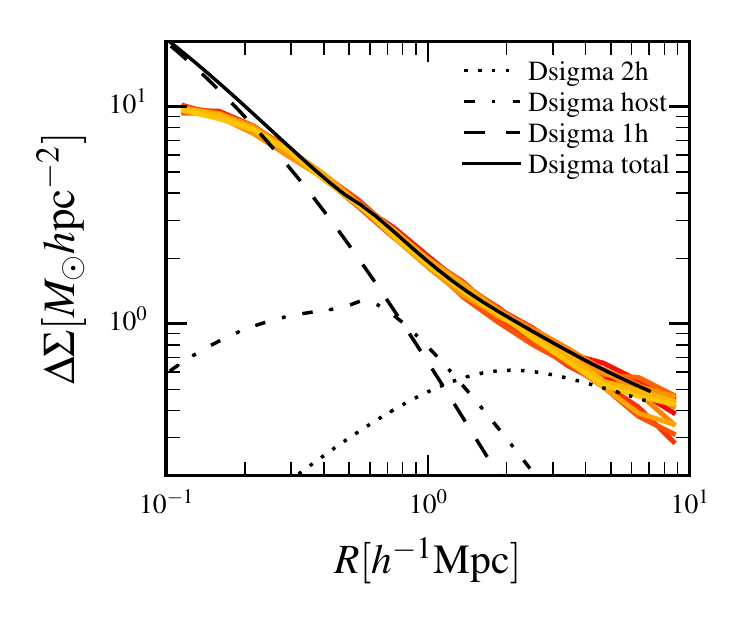}
	\caption{ Lensing signal produced by haloes with $M_{\rm 1h} \in [1\times10^{12}; 2\times10^{12}]h^{-1}M_{\odot}$, from the BigMultiDark simulation. The best-fit model gives a best-fit halo mass of $M_{\rm 1h} = (1.42 \pm 0.07)10^{12}h^{-1}M_{\odot}$ and a best-fit host mass of $M_{\rm host} = (2.20 \pm 0.20)10^{14}h^{-1}M_{\odot}$. The lack of signal at small scales ($R< 0.3h^{-1}\si{Mpc}$) is due to the resolution of the simulation.
          }
	\label{fig:lensing_M1-2e12}
\end{figure}

To verify the predictions of the model, we use the BigMultiDark simulated data. We compute the lensing signal for all the haloes with $1\times10^{12} < M_{\rm h}/h^{-1}M_{\odot} < 2\times10^{12}$ for the 15 lightcones. We fit the model to the mean lensing signal over the 15 computed, to obtain a best-fit halo mass of $M_{\rm 1h} = (1.42 \pm 0.07)10^{12}h^{-1}M_{\odot}$ and a best-fit host mass of $M_{\rm host} = (2.20 \pm 0.20)10^{14}h^{-1}M_{\odot}$, which is consistent with the input. In this case, as our lens sample is a mixture of central and satellite galaxies, the satellite fraction is not equal to one, and according to the data we fix $f_{\rm sat} = 0.09$. The lensing signal and best-fit model are shown in figure \ref{fig:lensing_M1-2e12}.

The model gives a good description of the simulated signal, except at small scales ($R < 0.3 h^{-1}\si{Mpc}$), where the limited resolution of the simulation damps the signal.

	\subsection{Contamination by non-satellite galaxies}
	\label{sec:contamination}
	
Even if we select only the satellite galaxies in the redMaPPer catalogue with a probability of membership higher than 0.8, our samples can still contain line-of-sight galaxies which are not member of the clusters. We test their influence on the lensing signal using the BigMultiDark simulations.

We use as hosts the haloes with $M_{\rm h} > 10^{14}h^{-1}M_{\odot}$ in the redshift range $0.2 < z_{\rm host} < 0.55$, and select the satellites in the same cluster-centric distance bins as in our measurement: $0.1h^{-1}\si{Mpc} < R_{\rm s} < 0.55h^{-1}\si{Mpc} $ and $0.55 h^{-1}\si{Mpc} < R_{\rm s} < 1 h^{-1}\si{Mpc} $. We make the following redshift selections to obtain one sample with only real satellites, and one sample contaminated by galaxies in the line-of-sight:
\begin{itemize}
	\item the pure satellite sample: we select galaxies with $|z_{\rm gal} - z_{\rm host}| \leq 0.0007$. In the redshift range $0.2-0.55$, this corresponds to $|d_{\rm gal} - d_{\rm host}| \leq 1.5 h^{-1}\si{Mpc} \sim R_{\rm vir}^{\rm host}$.
	\item the contaminated sample: here we have $|z_{\rm gal} - z_{\rm host}| \leq 0.05$. This gives a contamination rate of 37\% in the inner radial bin and 65\% in the outer radial bin.
\end{itemize}

\begin{figure}
	\centering
	\includegraphics{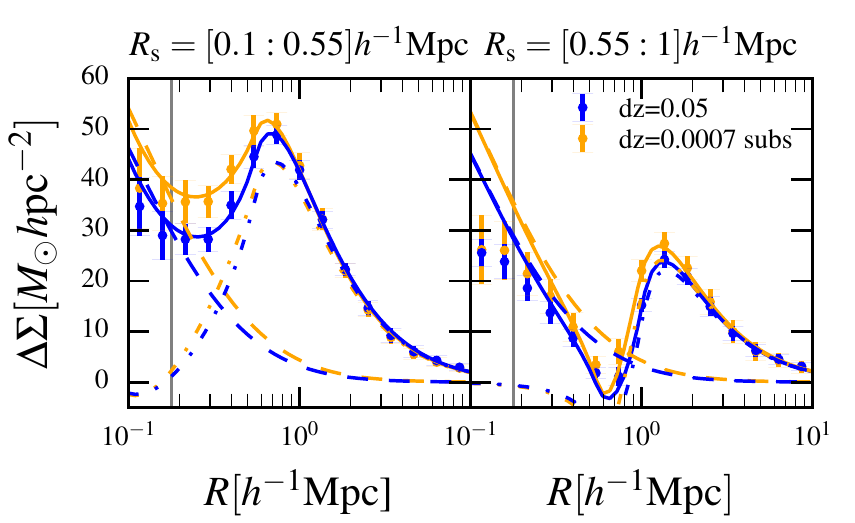}
	\caption{Lensing signal of satellite galaxies in the BDM simulations, with and without contaminants. The best-fit masses are presented in table \ref{tab:contaminants}.
         }
	\label{fig:contaminants}
\end{figure}

We show in figure \ref{fig:contaminants} the lensing signal computed in both radial bins, for the pure and the contaminated samples. In both radial bins, when contaminants are added, the amplitude of the signal decreases at small scales, inducing a drop in the subhalo mass: for the satellites in the inner part (respectively the outer part) of clusters, the best-fit satellite mass goes from $13.02 \pm 0.03$ (respectively $13.00 \pm 0.04$) for the pure sample to $12.88 \pm 0.03$ (respectively $12.86 \pm 0.04$) with contaminants. All the best-fit masses are summarized in table \ref{tab:contaminants}. To take into account this drop, we should increase our measurements by 1\%, which does not change our conclusions. 

\begin{table}
\centering
\begin{tabular}{c c c c}
$z_{\rm g} - z_{\rm h}$	&	$R_{\rm s}$		&	$\log(M_{\rm sat})$			&	$\log(M_{\rm host})$			\\
\hline
\multirow{2}{*}{0.0007}	&	0.1-0.55			&	$13.02 \pm 0.03$			&	$14.25 \pm 0.01$			\\
					&	0.55-1			&	$13.00 \pm 0.04$			&	$14.31 \pm 0.03 $			\\
\hline
\multirow{2}{*}{0.05}		&	0.1-0.55			&	$12.88 \pm 0.03$			&	$14.28 \pm 0.01$			\\
					&	0.55-1			&	$12.86 \pm 0.04$			&	$14.29 \pm 0.02$			\\
		
\hline
\end{tabular}
\caption{Best-fit masses for a sample of pure satellite galaxies and a sample contaminated by line-of-sight effects. The distances are expressed in $h^{-1}\si{Mpc}$ and the masses in $h^{-1}M_{\odot}$.}
\label{tab:contaminants}
\end{table}

In addition to line-of-sight projections, the miscentring of clusters could be another source of contaminations \citep{george2012}: if the galaxy we use as centre of the cluster is not its true centre, we can identify outer part satellites as inner ones and vice-versa. To verify the importance of this effect on the signal, we make the measurements again, selecting this time only clusters with a probability of having their true centre identified higher than 0.5. This selection has no significant impact on the signal, other than increasing the error bars due to the decrease in the number of lenses.

\section{Systematic error}
\label{sec:syst_err}

	\subsection{Consistency of the three source catalogues}
	\label{sec:cat_comp}

\begin{figure}
	\centering
	\includegraphics{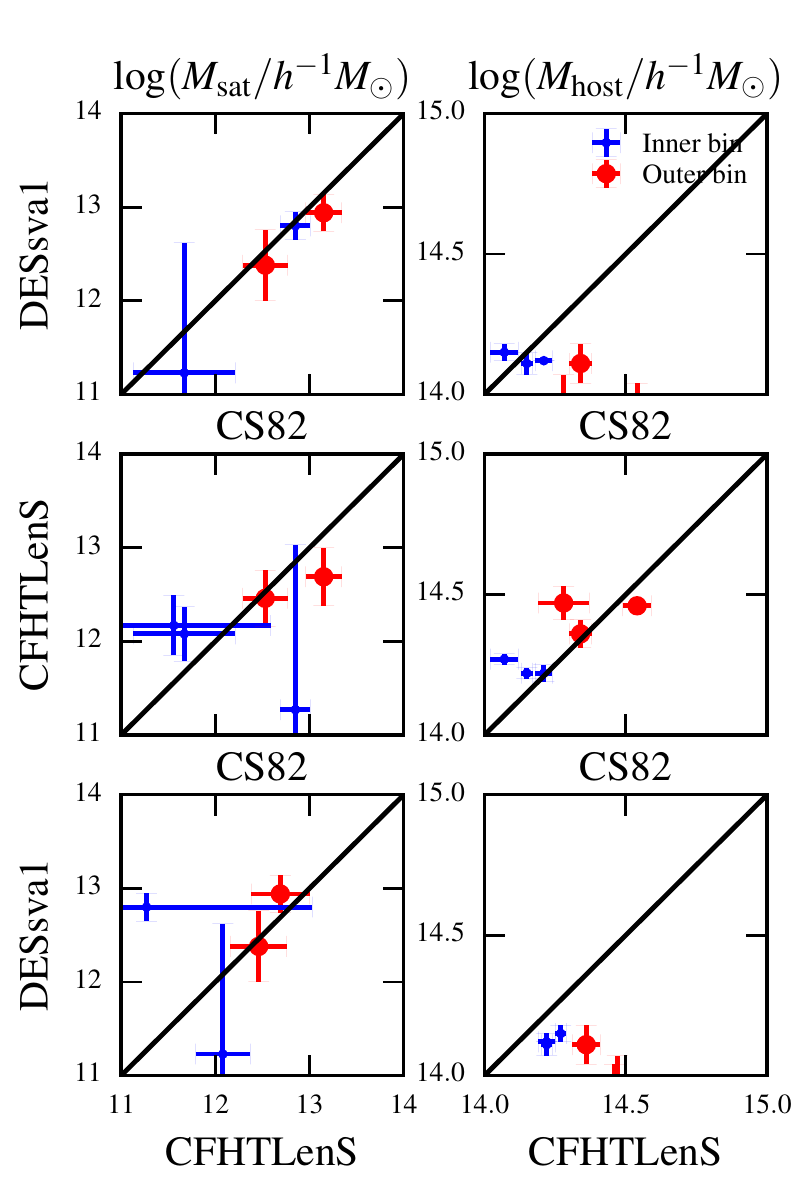}
	\caption{Comparison of the mass measurements using the CS82 and CFHTLenS and DES-SV catalogues. The left column shows the satellite mass and the right panel the host mass. We plot the $x=y$ line to guide the eye.
          }
	\label{fig:cs82_vs_cfhtlens_vs_dessva1}
\end{figure}

We measured the haloes of satellite galaxies in three bins in stellar mass, and in two ranges in distance to the centre of the cluster, using shear catalogues from the CS82, CFHTLenS and DES-SV surveys.
Each of these surveys have different photometric measurements, as well as different estimators of the photo-z, stellar mass and shape. To ensure that no bias is introduced by either of the catalogues, we repeat the same measurements using each of the catalogues separately. We compare the obtained masses in figure \ref{fig:cs82_vs_cfhtlens_vs_dessva1}, and although the uncertainties are high, the different measurements of the satellite mass appear to be consistent.

	\subsection{Redshift accuracy}
	\label{sec:red_comp}
	
Different cuts are possible in the source catalogues to have photometric redshifts as clean as possible. As shown in \citet{heymans2012}, keeping only galaxies with $0.2 < z < 1.3$ ensures a relatively accurate photometric redshift, with a typical uncertainty of $\sigma_z \sim 0.04(1 + z)$. 
In addition, cutting objects with \textsc{ODDS} $< 0.5$ secures even further the accuracy of redshifts.
We test both cuts on the CS82 and CFHTLenS catalogues: for CS82 the initial catalogue contains 4,381,917 objects, and keeps $87\%$ with $0.2 < z < 1.3$, and $64\%$ with \textsc{ODDS} $> 0.5$. For CFHTLenS, the uncut catalogue contains 7,511,368 objects, going down to $75\%$ with $0.2 < z < 1.3$ and $89\%$ with \textsc{ODDS} $> 0.5$.

\begin{figure}
	\centering
	\includegraphics{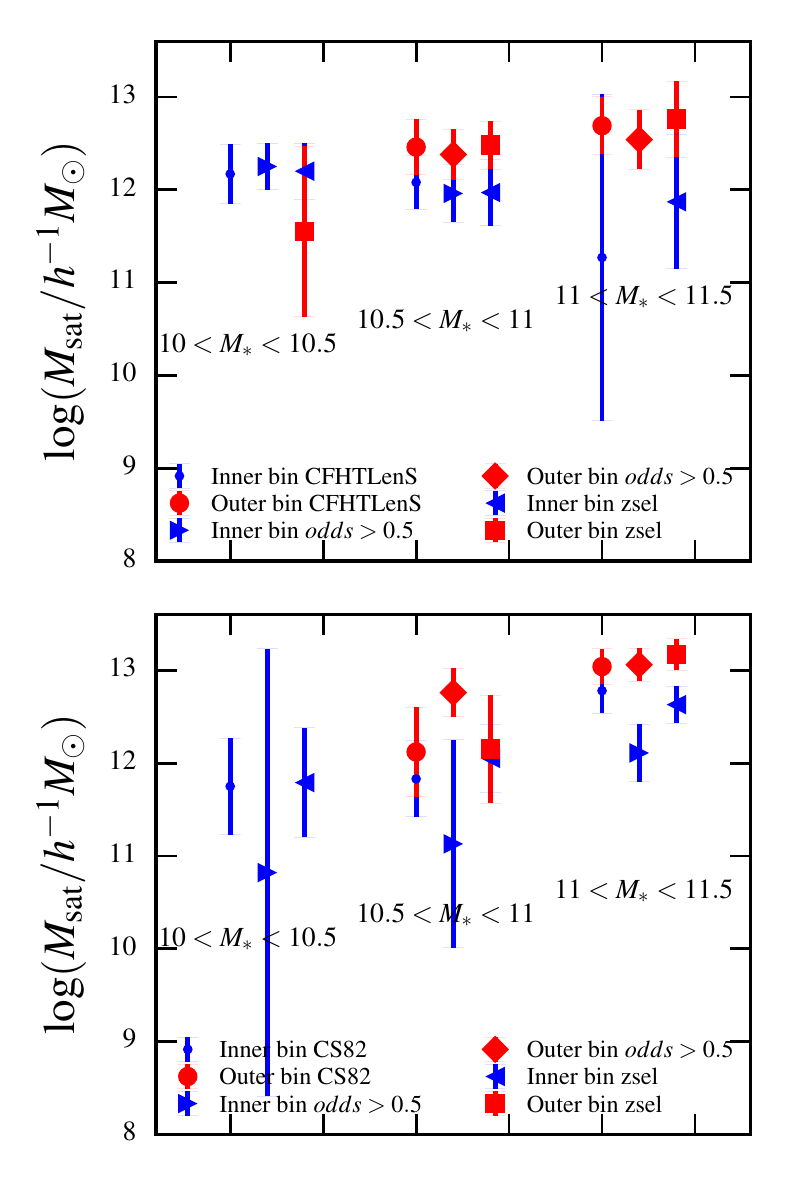}
	\caption{\textit{Top panel:} best-fit results for the satellite masses computed using only the CFHTLenS source catalogue. The blue symbols are the masses for satellites from the inner part of the clusters, and the red symbols for satellites from the outer part of the clusters. We compute the masses using three different source catalogues: all the sources, sources selected with \textsc{ODDS} $> 0.5$ and sources with $0.2 < z < 1.3$. \textit{Bottom panel:} same but for the CS82 source catalogue.
          }
	\label{fig:cfhtlens_cs82_comp}
\end{figure}

 We show in figure \ref{fig:cfhtlens_cs82_comp}  the best-fit subhalo masses computed using the different cuts. We find that the different results are very similar, and we choose to use the cut $0.2 < z < 1.3$ which appears to give the best contraints.

\bibliographystyle{mn2e}
\bibliography{/Users/aniemiec/Biblio/lensing.bib}

\label{lastpage}
\end{document}